\documentclass[12pt]{article}
\usepackage{amsfonts,amsmath,amssymb,epsf}
\usepackage{amsmath,braket}
\usepackage{graphicx,hyperref,color}
\usepackage{cite}
\hypersetup{
    bookmarks=true,         
    unicode=false,          
    pdftoolbar=true,        
    pdfmenubar=true,        
    linktocpage=true,       
    pdffitwindow=false,     
    pdfstartview={FitH},    
    pdfnewwindow=true,      
    colorlinks=true,       
    linkcolor=blue,          
    citecolor=blue,        
    filecolor=blue,      
    urlcolor=blue           
}
\numberwithin{equation}{section}									

\def\d#1{\,{\rm d}#1}

\newcommand{\de}{\partial}
\newcommand{\be}{\begin{equation}}
\newcommand{\ba}{\begin{eqnarray}}
\newcommand{\ea}{\end{eqnarray}}
\newcommand{\ee}{\end{equation}}

\newcommand{\f}{\frac}
\newcommand{\s}{\sqrt}
\newcommand{\vp}{\varphi}

\newcommand{\ti}{\tilde}
\newcommand{\ap}{\alpha}

\newcommand{\ddd}{\cdot\cdot\cdot}
\newcommand{\no}{\nonumber \\}
\newcommand{\la}{\langle}
\newcommand{\lb}{\rangle}
\newcommand{\bea}{\begin{eqnarray}}
\newcommand{\eea}{\end{eqnarray}}
\newcommand{\bes}{\begin{equation*}}
\newcommand{\beas}{\begin{eqnarray*}}
\newcommand{\eeas}{\end{eqnarray*}}
\newcommand{\bas}{\begin{array*}}
\newcommand{\eas}{\end{array*}}
\newcommand{\ees}{\end{equation*}}
\newcommand{\nn}{\nonumber}
\newcommand{\p}{\partial}
\newcommand{\ep}{\epsilon}

\newcommand{\onshell}{\scriptsize\mbox{on-shell}}
\let\a=\alpha \let\b=\beta \let\c=\chi \let\d=\delta \let\e=\epsilon  \let\g=\gamma \let\h=\eta \let\k=\kappa \let\l=\lambda \let\m=\mu \let\n=\nu
 \let\p=\phi \let\r=\rho 
\let\t=\tau   \let\vp=\varphi   
 \let\D=\Delta \let\G=\Gamma \let\L=\Lambda   \let\S=\Sigma    
\def\nn{\nonumber}
\def\inf{\infty}

\def\pa{\partial}

\begin{document}

\begin{titlepage}
\thispagestyle{empty}

\vspace*{-2cm}
\begin{flushright}
YITP-22-14
\\
IPMU22-0002
\\
\end{flushright}

\bigskip

\begin{center}
\noindent{{\Large \textbf{BCFT and Islands in Two Dimensions}}}\\
\vspace{2cm}

\quad Kenta Suzuki$^a$ \ and \ Tadashi Takayanagi$^{a,b,c}$
\vspace{1cm}\\

{\it $^a$Center for Gravitational Physics,\\
Yukawa Institute for Theoretical Physics,
Kyoto University, \\
Kitashirakawa Oiwakecho, Sakyo-ku, Kyoto 606-8502, Japan}\\
\vspace{1mm}
{\it $^b$Inamori Research Institute for Science,\\
620 Suiginya-cho, Shimogyo-ku,
Kyoto 600-8411 Japan}\\
\vspace{1mm}
{\it $^{c}$Kavli Institute for the Physics and Mathematics
 of the Universe (WPI),\\
University of Tokyo, Kashiwa, Chiba 277-8582, Japan}\\

\bigskip \bigskip
\vskip 2em
\end{center}

\begin{abstract}

By combining the AdS/BCFT correspondence and the brane world holography,
we expect an equivalence relation between a boundary conformal field theory (BCFT) and a gravitational system coupled to a CFT.
However, it still remains unclear how the boundary condition of the BCFT is translated in the gravitational system.
We examine this duality relation in a two-dimensional setup by looking at the computation of entanglement entropy and energy flux conservation. We also identify the two-dimensional gravity which is dual to the boundary dynamics of a BCFT. Moreover, we show that by considering a gravity solution with scalar fields turned on, 
we can reproduce one point functions correctly in the AdS/BCFT.

\end{abstract}

\end{titlepage}

\newpage

\tableofcontents

\section{Introduction}
\label{sec:introduction}

The lsland formula \cite{Penington:2019npb,Almheiri:2019psf,Almheiri:2019hni} has lead us new insights on how entanglement entropy in a CFT behaves when it is coupled to a gravitational theory. In particular, the lsland formula gives a remarkable explanation of the Page curve \cite{Page:1993wv, Page:2013dx} in black hole evaporation processes. Even though we can directly derive the Island formula in two dimensional gravity by taking into account the replica wormhole contributions \cite{Penington:2019kki,Almheiri:2019qdq}, we need to rely on indirect arguments to justify the Island formula in higher dimensions at present. One argument is to consider the holographic entanglement entropy formula \cite{Ryu:2006bv,Ryu:2006ef,Hubeny:2007xt} with quantum corrections \cite{Faulkner:2013ana,Engelhardt:2014gca}. This formula has originally been considered to be applicable to asymptotically AdS backgrounds, assuming the AdS/CFT \cite{Maldacena:1997re}. In principle, however we can straightforwardly generalize the holographic entanglement entropy formula to any gravitational backgrounds such as asymptotically flat spacetimes, though their dual field theories are not clear. This formally leads to the Island formula.
\footnote{
Further discussion on the lsland formula are given for example in \cite{Akers:2019nfi, Almheiri:2019yqk, Almheiri:2019psy, Rozali:2019day, Chen:2019uhq, Bousso:2019ykv, Liu:2020gnp, Balasubramanian:2020hfs, Verlinde:2020upt, Chen:2020wiq, Gautason:2020tmk, Anegawa:2020ezn, Hashimoto:2020cas, Hartman:2020swn, Hollowood:2020cou, Krishnan:2020oun, Alishahiha:2020qza, Geng:2020qvw, Chen:2020uac, Chen:2020hmv, Hernandez:2020nem, Li:2020ceg, Chandrasekaran:2020qtn, Bak:2020enw, Bousso:2020kmy, Dong:2020uxp, Engelhardt:2020qpv, Chen:2020jvn, Chen:2020tes, Hartman:2020khs, Balasubramanian:2020xqf, Balasubramanian:2020coy, Sybesma:2020fxg, Ling:2020laa, Harlow:2020bee, Chen:2020ojn, Goto:2020wnk, Hsin:2020mfa, Akal:2020twv}.
}

\begin{figure}
  \centering
  \includegraphics[width=7cm]{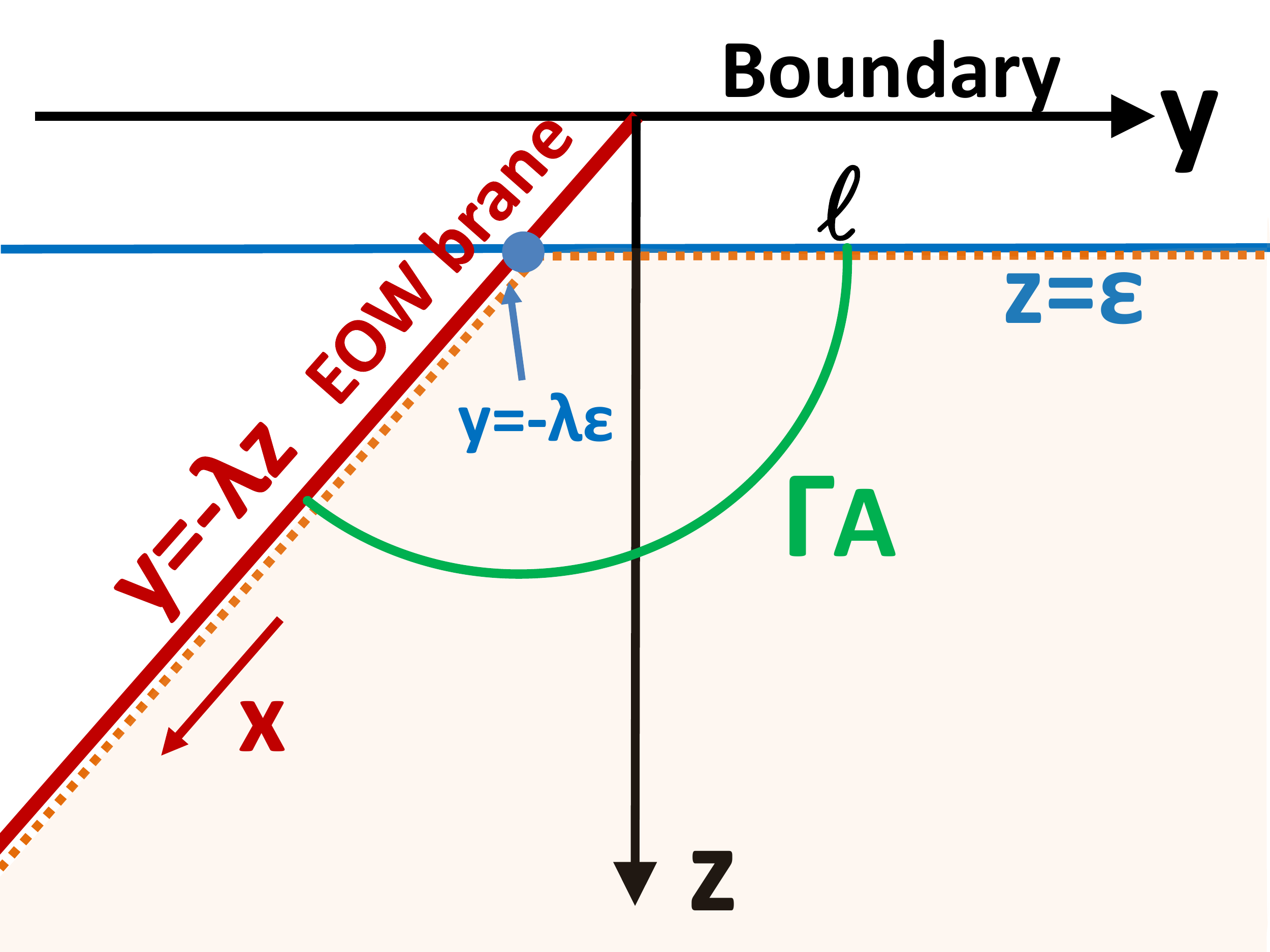}
  \caption{A sketch of AdS/BCFT setup.}
\label{fig:AdSBCFT}
\end{figure}

Another way to derive the Island formula is to employ the double holography argument \cite{Almheiri:2019hni}. A bulk dual of a CFT on R$^d$ coupled to a quantum gravity on AdS$_d$ can be described by inserting an end of the world-brane (EOW brane) in an $d+1$ dimensional bulk AdS by applying the brane-world holography \cite{Randall:1999ee,Randall:1999vf,Gubser:1999vj,Karch:2000ct},
where the gravity on the $d$ dimensional brane arises as an induced gravity (refer to Figure~\ref{fig:AdSBCFT}). 
Interestingly, the same gravity setup appears when we consider a gravity dual of a boundary conformal field theory (BCFT), so called AdS/BCFT  \cite{Karch:2000gx,Takayanagi:2011zk,Fujita:2011fp,Nozaki:2012qd}.
\footnote{
The AdS/BCFT has been applied to many problems.  This includes the studied of renormalization group flow \cite{Gutperle:2012hy,Estes:2014hka,Kobayashi:2018lil,Sato:2020upl},  
applications to models in condensed matter physics or statistical mechanics \cite{Fujita:2012fp,Ugajin:2013xxa,Erdmenger:2015xpq,Seminara:2017hhh,Seminara:2018pmr,Hikida:2018khg,Shimaji:2018czt,Caputa:2019avh,Mezei:2019zyt}, and to computational complexity \cite{Chapman:2018bqj,Sato:2019kik,Braccia:2019xxi,Sato:2021ftf}. Refer also to \cite{Chiodaroli:2011nr,Chiodaroli:2012vc,Karch:2020iit,Bachas:2020yxv,Simidzija:2020ukv,Ooguri:2020sua,Raamsdonk:2020tin} for string theory embeddings, to \cite{Cooper:2018cmb,Antonini:2019qkt,VanRaamsdonk:2020tlr,VanRaamsdonk:2021qgv} for application to cosmological models and to \cite{Akal:2020wfl,Miao:2020oey} for higher codimension holography.} Thus, there are two different interpretations of an identical gravity dual with an EOW brane. Both of them have the same rule to calculate the holographic entanglement entropy, namely the minimal area surface, whose area gives the entanglement entropy, can end on the EOW brane. 

\begin{figure}[t!]
  \centering
  \includegraphics[width=7cm]{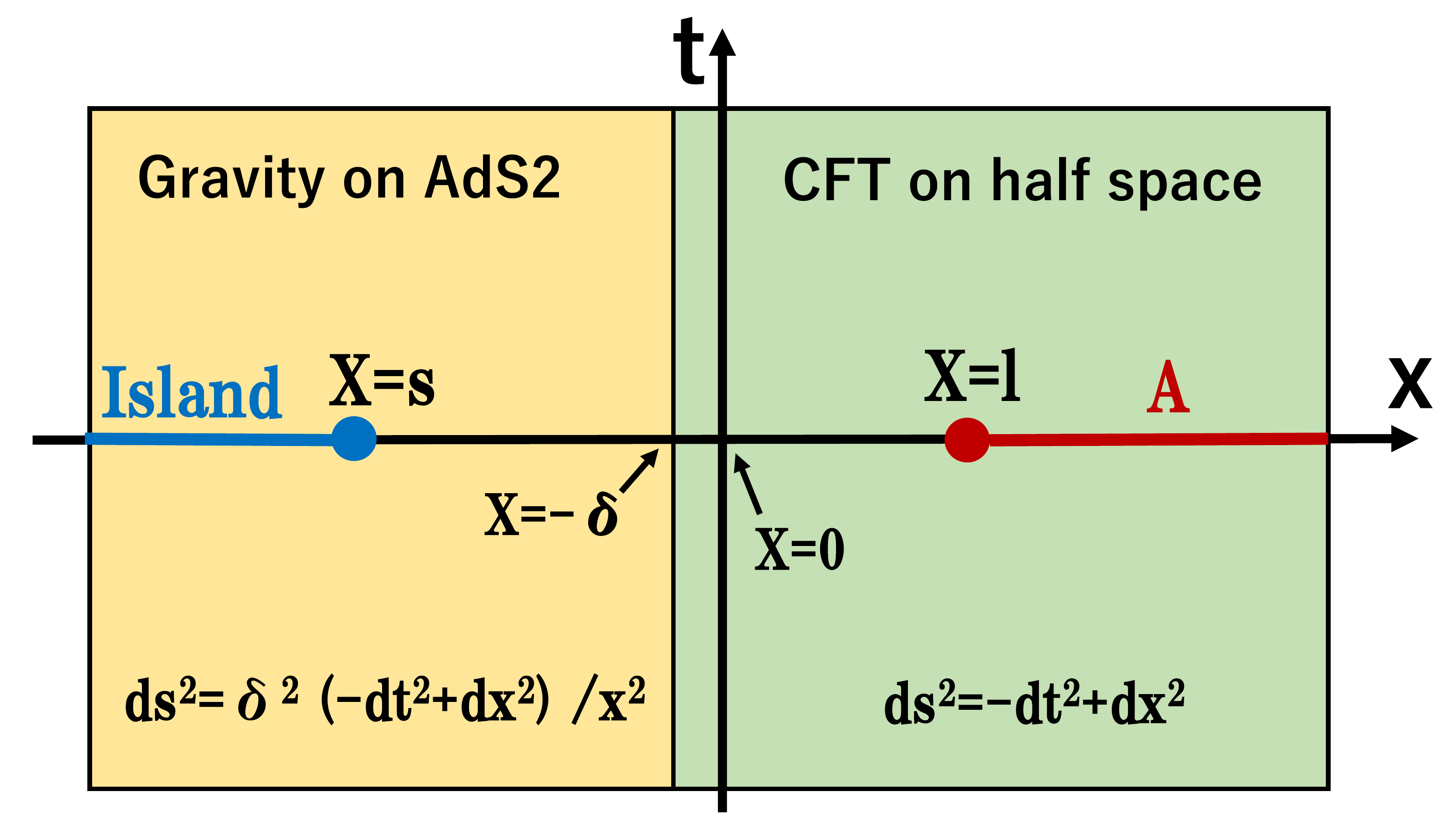}
 \includegraphics[width=7cm]{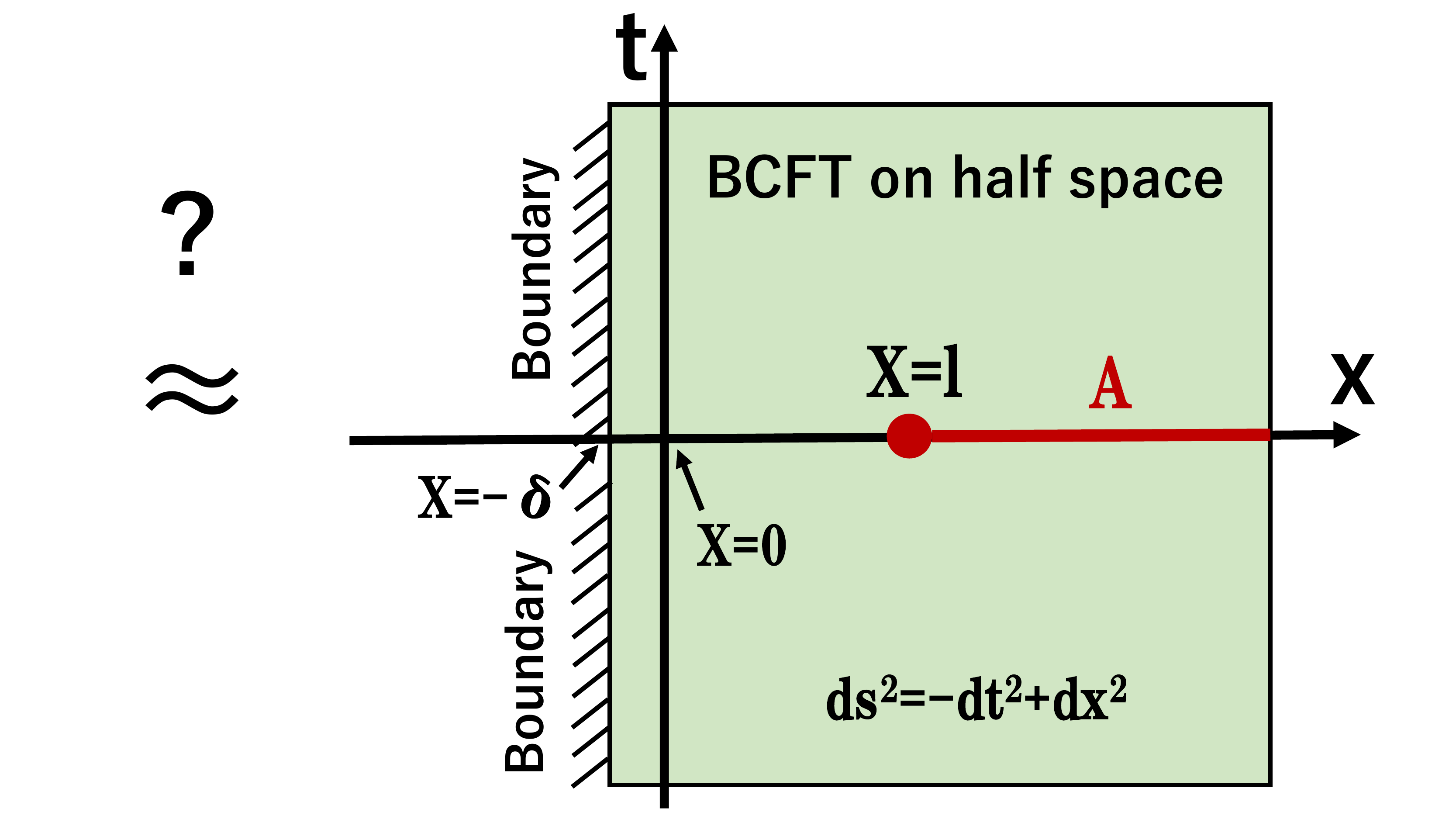}
  \caption{A sketch of setup of calculating entanglement entropy when we couple a CFT on a half line to a gravity on AdS$_2$ (left) and an equivalent setup in BCFT (right).}
\label{fig:Island}
\end{figure}

From this holographic argument one may wonder if the two boundary theories: (i) a CFT coupled to a gravity and (ii) a BCFT, are equivalent as depicted in Figure~\ref{fig:Island}. The purpose of this paper is to argue that this is indeed the case by presenting evidences of the equivalence by focusing on two dimensional (2d) CFTs $d=2$. We call this equivalence the Island/BCFT correspondence.
We will show not only that the calculation of entanglement entropy matches between the two, but also that the energy flux reflection in (ii) the BCFT, can be obtained from (i) the CFT coupled to a gravity. We will also identify the 2d gravity realized on the EOW brane in the case of pure AdS$_3$ gravity dual.
Finally, we also consider bulk one point functions, which are a part of essential information of a given BCFT. We will show that to obtain non-vanishing bulk one-point functions in BCFTs we need to modify the prescription of \cite{Takayanagi:2011zk,Fujita:2011fp} such that we turn on non-trivial background matter fields. The Neumann boundary condition of matter fields imposed on the EOW brane induces this non-trivial matter field background.

The paper is organized as follows. In section~\ref{sec:island/bcft}, we calculate the entanglement entropy in a 2d CFT coupled to a 2d gravity and show that it agrees with the result form a BCFT. From this we argue the Island/BCFT correspondence.
In section~\ref{sec:gravity dual}, we calculate the entanglement entropy of a BCFT from the gravity dual by using the AdS/BCFT and compare the result with that obtained in section~\ref{sec:island/bcft}. In section~\ref{sec:2d gravity}, we derive the 2d gravity dynamics induced on the EOW brane from both holographic and field theoretic analysis in the case of pure AdS$_3$ gravity dual. We identify the 2d induced gravity dual to the boundary dynamics of BCFTs via the Island/BCFT correspondence.
In section \ref{sec:replica wormholes}, we explicitly show the boundary condition of energy stress tensor in a BCFT matches with that derived from a 2d CFT coupled to a 2d gravity. Moreover we derive the Island formula used in section~\ref{sec:island/bcft} from the replica wormhole calculation. In section~\ref{sec:one point function}, we show that by turning on bulk matter fields, we can reproduce one point functions in a BCFT from the AdS/BCFT. For a bulk one point function of an exactly marginal operator, we can analytically construct such a background by using the known Janus solution. For more general operators, 
we numerically find the gravity dual backgrounds with a massive scalar field turned on. In section~\ref{sec:conclusions}, we summarize our results and discuss future problems.
Since we use various coordinates in this paper, we summarize our notation in appendix~\ref{app:notation}.
In appendix~\ref{app:onshell action}, we present a naive discussion about on-shell action of the induced gravity we discussed in section~\ref{sec:2d gravity}.
In appendix~\ref{app:adm energy}, we summarize the ADM energy in JT gravity, which is originally discussed in \cite{Maldacena:2016upp}
and in appendix~\ref{app:replica wormholes}, we review the replica wormholes in JT gravity coupled to a conformal matter, following the discussion of \cite{Almheiri:2019qdq}.

\section{Island/BCFT Correspondence}
\label{sec:island/bcft}

We consider a two dimensional (2d) CFT on a half line $x\geq -\delta$ coupled to a two dimensional gravity on AdS$_2$ as depicted in the left of Figure~\ref{fig:Island}.
We assume $\delta$ to be an infinitesimally small positive constant so that the spacetime of gravity includes an asymptotically AdS$_2$ region.
We choose the coordinate $(t,x)$ such that the metric is given by 
\footnote{We summarize our notation in appendix~\ref{app:notation}.}
\ba
	ds_{(2)}^2=e^{2\phi(x)}(-dt^2+dx^2),  \label{metre}
\ea
where the Weyl factor takes the form
\ba
&& e^{2\phi(x)}=1 \qquad \ \ \mbox{for} \ \  x\geq -\delta , \no
&& e^{2\phi(x)}=\frac{\delta^2}{x^2} \qquad \mbox{for} \ \  x< -\delta . \label{methet}
\ea

Now we choose the subsystem $A$ in the CFT$_2$ in the right of Figure~\ref{fig:Island} to be the half line $x\geq l$ and compute the entanglement entropy $S_A$.
Since we have gravity in the left half, we need to employ the Island prescription \cite{Penington:2019npb,Almheiri:2019psf,Almheiri:2019hni} to calculate $S_A$.
We choose the island region $I_s$ to be an semi-infinite line 
$x<s$, where we assume $s<-\delta$. The island formula tells us that to calculate $S_A$ we need to evaluate the field theory entanglement entropy for the region 
$A\cup I_s$ plus a gravity area contribution $S_{gravity}$ of the boundary of the Island and then we can determine $s$ by minimizing the total contribution. In our current setup, $S_A$ can be found by minimizing 
\ba
S_A=\frac{c}{3}\log \left(\frac{l-s}{\ep}\right) +\frac{c}{6}\log \left(\frac{\delta}{-s}\right)+S_{gravity},
\ea 
with respect to $s$, where $c$ is the central charge of the CFT$_2$ and $\ep$ is the UV cut off. The first term is the standard entanglement entropy in two dimensional CFT \cite{Holzhey:1994we} and the second term, which is identical to $\frac{c}{6}\phi(x=s)$,
arises due to the Weyl rescaling in order to take into account the non-trivial metric on AdS$_2$. The final term $S_{gravity}$ denotes all of the gravitational contributions.

We would like to consider an induced gravity where the full gravity action is produced by integrating the matter CFT degrees of freedom, though the metric (i.e. $\phi$) is dynamical. In this case we can ignore $S_{gravity}$, as the all induced gravity contributions are included in the CFT parts. In this case, the minimization of $S_A$ is achieved at 
\ba
s=-l.  \label{mirror}
\ea
The resulting entanglement entropy reads
\ba
S_A=\frac{c}{6}\log \left(\frac{2l}{\ep}\right)+\frac{c}{6}\log\left(\frac{2\delta}{\ep}\right). 
\label{EEIS}
\ea

Interestingly if we assume $\delta$ is infinitesimally small such that $\delta=O(\ep)$, the above expression (\ref{EEIS}) 
takes the same form as that for a 2d BCFT \cite{Calabrese:2004eu}, as sketched in the right of Figure~\ref{fig:Island}. In this relation, which may be called as Island/BCFT correspondence, the second term in the entanglement entropy (\ref{EEIS}) can be identified with the boundary entropy $S_{bdy}$ \cite{Affleck:1991tk}.
The end point $x=-l$ of the Island can be interpreted as the mirror image of the endpoint of the subsystem $A$, 
which indeed arises when we perform the replica calculation of $S_A$ in the BCFT using the twist operators \cite{Calabrese:2004eu}.
To get the location of this mirrored point (\ref{mirror}), the AdS$_2$ geometry plays a crucial role. 
However, up to now, we did not use holography at all.

\section{Entanglement Entropy in AdS/BCFT}
\label{sec:gravity dual}

We can give a gravity dual of the previous correspondence between the entanglement entropy in the Island prescription and that in BCFT's by employing the AdS/BCFT correspondence \cite{Takayanagi:2011zk,Fujita:2011fp}.  We focus on the case where the boundary is a straight line. For non-trivial shapes of
boundaries refer to \cite{Nozaki:2012qd}.

The total action of the gravity dual in AdS/BCFT is given by
	\begin{align}
		I_{\rm total} \, = \, I_{\rm bulk} \, + \, I_{\rm brane} \, + \, I_{\rm bdy} \, ,
	\end{align}
where
	\begin{align}
		I_{\rm bulk} \, &= \, - \, \frac{1}{16\pi G_N} \int_M d^3x \, \sqrt{-g}(R- 2\L) \, , \\
		I_{\rm brane} \, &= \, - \, \frac{1}{8\pi G_N} \int_Q d^2x \, \sqrt{-h}(K_Q - T) \, , \\
		I_{\rm bdy} \, &= \, - \, \frac{1}{8\pi G_N} \int_\S d^2x \, \sqrt{-\g} K_\S \, ,
	\end{align}
where $R$ is Ricci scalar and $K$ is the trace of extrinsic curvature $K_{ij}$.
We denote the induced metric on the brane by $h_{ij}$ and that on the asymptotic boundary by $\g_{ij}$.
By choosing the AdS radius to be unit i.e. $\Lambda=-1$, in the Poincare AdS$_3$,
\ba
ds_{\rm AdS_3}^2=\frac{-dt^2+dy^2+dz^2}{z^2},
\ea
we introduce the end of the world brane $Q$ (EOW brane)  as the two dimensional plane specified by
\ba
y=-\lambda z,
\ea
where $\lambda>0$ is a parameter related to the tension ${\cal{T}}$ of the brane via
\ba
\lambda=\frac{\cal T}{\s{1-{\cal T}^2}}.
\label{lambda}
\ea
The region given by $z>\ep$ and $y>-\lambda z$ provides a gravity dual of a BCFT defined as a two dimensional CFT on the half space $y>-\lambda \ep$ via the AdS/BCFT
\cite{Takayanagi:2011zk,Fujita:2011fp}. 
This is depicted in Figure~\ref{fig:AdSBCFT}.
The metric of EOW-brane is written as
that of the AdS$_2$:
\ba
ds_{(2)}^2 \, = \, \frac{(1+\lambda^2)dz^2-dt^2}{z^2} \, = \, (1+\lambda^2)\frac{dx^2-dt^2}{x^2},  \label{adstw}
\ea
where we introduced the coordinate $x$ along the EOW brane by $x^2=y^2+z^2$ so that the metric takes the form of (\ref{metre}). 
We can equivalently employ the hyperbolic slice of AdS$_3$
\ba
ds^2=d\rho^2+\cosh^2\rho\left(\frac{-dt^2+d\eta^2}{\eta^2}\right),
\label{metric ansatz}
\ea
where the new coordinates $\eta$ and $\rho$ are defined by 
\ba
z \, = \, \frac{\eta}{\cosh\rho} \, , \qquad y \, = \, \eta\tanh\rho \, .
\label{z&y}
\ea
Then the gravity dual of the BCFT is given by the region $-\rho_*\leq \rho\leq\infty$ and 
the brane $Q$ is located at a constant $\r = \r_*$ and its value is determined by the brane tension ${\cal{T}}$ as
	\begin{align}
	    {\cal{T}} \, = \, \tanh \r_* \, .
	\end{align}

For the boundary BCFT$_2$, we consider the vacuum state (which is a pure state) and thus we have $S_A = S_{\bar{A}}$, where $\bar{A}$ is the complement of the subgreion $A$ defined by $-\l\e < y < l$.
The holographic entanglement entropy \cite{Ryu:2006bv, Ryu:2006ef} of a subregion $A$ is computed in terms of the area of the codimension two minimal surface
(called $\G_A$) anchored at $A$ as
	\begin{align}
		S_A \, = \, \frac{\textrm{Area}(\G_A)}{4G_N} \, .
	\end{align}
In the setup of of the AdS$_3$/BCFT$_2$, the minimal surface $\G_{\bar{A}}$ is given by the spacial geodesic $\Gamma_A$ as depicted in Figure~\ref{fig:AdSBCFT}.
Since $\Gamma_A$ is the arc with the radius $l$, the holographic calculation leads to 
	\begin{align}
		S_A \, = \, \frac{1}{4G_N} \int_{-\r_{\inf}}^{\r_*} d\r \, = \, \frac{\r_\inf + \r_*}{4G_N} \, = \, \frac{c}{6} \, (\r_\inf + \r_*) \, ,  
	\end{align}
where we used the dictionary between the AdS$_3$ radius $L_{\textrm{AdS}_3}$ and the central charge $c$ of a CFT$_2$ \cite{Brown:1986nw}
	\begin{align}
		c \, = \, \frac{3L_{\textrm{AdS}_3}}{2G_N} \, ,
	\end{align}
and we set $L_{\textrm{AdS}_3}=1$.
Using the relation between the coordinates (\ref{z&y}), we have $\r_{\inf} = \textrm{arccosh}(l/\e) \approx \log(2l/\e)$ and $\r_* = \textrm{arctanh}(\cal{T})$.
Combining these results, we finally find 
	\begin{align}
        S_A \, &= \, \frac{c}{6}\log \left(\frac{2l}{\ep}\right)+\frac{c}{6}\log\s{\frac{1+{\cal T}}{1-{\cal T}}}\no
        \, &= \, \frac{c}{6}\log\left(\frac{2l}{\ep}\right)+\frac{c}{6} \log(\lambda+\s{1+\lambda^2}) \, .  
    \label{EEISA}
	\end{align}
This is the prediction by AdS/BCFT for the entanglement entropy in the dual BCFT.

Since we impose the Neumann boundary condition on the EOW brane, there is another interpretation of our setup, namely 
a CFT on the half space coupled to a gravity on the AdS$_2$ (left side in Figure~\ref{fig:Island}).
By comparing this setup with the previous Island setup, we can identify (from $x^2=y^2+z^2$)
\ba
\delta=\s{1+\lambda^2} \, \ep,
\ea
which leads to the entanglement entropy (\ref{EEIS}) in the Island prescription, assuming an induced gravity, evaluated as 
\ba
S_A=\frac{c}{6}\log \left(\frac{2l}{\ep}\right)+\frac{c}{6}\log\left(2\s{1+\lambda^2}\right).  \label{EEISG}
\ea
When $\lambda$ is large, this nicely agrees with the AdS/BCFT result (\ref{EEISA}). This provides another support for the Island/BCFT correspondence. On the other hand, if $\lambda$ is not large, we expect the gravity on the EOW brane is highly quantum and we cannot 
trust the 2d gravity analysis using the classical geometry (\ref{adstw}), for which we may need more care analysis when we relate $\delta$ to $\ep$.

\section{2d Induced Gravity and Gravity Dual}
\label{sec:2d gravity}

Next we would like to identify the gravitational theory realized on the AdS$_2$ (EOW brane) by directly analyzing the AdS$_3$/BCFT$_2$ setup of Figure~\ref{fig:AdSBCFT}. In other words, this 2d gravity is supposed to be equivalent to the boundary dynamics of a BCFT via the Island/BCFT correspondence. For this, we will work with a 2d Euclidean space. 

Via the holography, the 2d effective gravity can be found as follows. We impose the Dirichlet boundary condition on the asymptotic AdS$_3$ region $\Sigma$ and solve the bulk Einstein equation for a fixed metric 
$ds^2=h_{ij}dx^i dx^j$ on the EOW brane $Q$. Then we find 
the on-shell Euclidean action $I_G[h]$ defined by
\ba
I_G \, = \, \frac{1}{16\pi G_N}\int_M \s{g}(R-2\Lambda) \, + \, \frac{1}{8\pi G_N}\int_\Sigma \s{\g}K \, + \, \frac{1}{8\pi G_N}\int_Q \s{h}(K-T).
\ea
Moreover, if we impose the saddle point equation  $\frac{\delta I_G[h]}{\delta h_{ij}}=0$ under the variation of the metric $h_{ij}$ on $Q$, then this gives the Neumann boundary condition 
\ba
K_{ij}-K h_{ij}+Th_{ij}=0,  \label{neubdyc}
\ea
as is standard in the AdS/BCFT. By solving this Neumann boundary condition, we find that the metric of EOW brane $Q$ is that of AdS$_2$. Therefore, we can identify the effective 2d gravity action on $Q$ with the on-shell action  $I_G[h]$.

\subsection{Naive Dimensional Reduction Argument}
Before diving into details of the induced gravity (which we will discuss in the next subsection),
let us first consider a naive dimensional reduction argument.

Using the hyperbolic slice coordinates of AdS$_3$ (\ref{metric ansatz}), 
now we regard the 2d quantum gravity which appears in the brane-world on $Q$ is dual to the 3d bulk on the region $-\rho_*\leq \rho\leq 0$,
where $\rho_*>0$ depends on $\lambda$ by 
\ba
\lambda=\sinh\rho_*.
\ea
Under this interpretation we can find the effective two dimensional Newton constant $G^{(2)}_N$ in terms of the three dimensional Newton constant $G_N$ via a simple dimensional reduction as follows
\footnote{
Given the metric ansatz (\ref{metric ansatz}), the precise reduction of the three-dimentional Ricci scalar is given by
	\begin{align}
		\s{-g^{(3)}}R^{(3)} \, = \, \s{-g^{(2)}} \Big[ R^{(2)} - 2 \big(3\cosh^2\r - 1\big) \Big] \, .
	\end{align}
Since we are interested in the contribution to gravitational entropy, we neglect the shift due to the second term.
}
	\begin{align}
        \frac{1}{16\pi G^{(2)}_N}\int d^2x \s{-g^{(2)}}R^{(2)} \, &= \, \frac{1}{16\pi G_{N}}\int d^3x \s{-g^{(3)}}R^{(3)} \no
        &\approx \, \frac{1}{16\pi G_{N}}\int^0_{-\rho_*} d\rho \int d^2x \s{-g^{(2)}}R^{(2)}
	\end{align}
This leads to
\ba
\frac{1}{G^{(2)}_N}=\frac{\rho_*}{G_N}.
\ea
In the two dimensional gravity picture, we have the AdS$_2$ entropy given by 
\ba
\frac{1}{4G^{(2)}_N}=\frac{\rho_*}{4G_N}=\frac{c}{6}\log(\lambda+\s{1+\lambda^2}). \label{gravtd}
\ea
Since in the present induced gravity treatment with the Newton constant $G^{(2)}_N$, the matter CFT was already integrated out, there is only gravitational contribution to the entanglement entropy. Thus the total entanglement entropy $S_A$ in the system where the induced gravity in the left half is coupled to a CFT in the right half, is obtained by the sum of (\ref{gravtd}) and the entanglement entropy in the right half CFT, estimated as $\sim\frac{c}{6}\log\frac{2l}{\ep}$.
This agrees with the entanglement entropy in the holographic BCFT result (\ref{EEISA}). 
In appendix~\ref{app:onshell action}, we also give a naive discussion for the on-shell action of the induced gravity.

\subsection{Deriving 2d Induced Gravity}

We argue that this 2d gravity on the EOW brane $Q$ is an induced gravity, where the original gravity action before the path-integration of the matter CFT, is simply given by a cosmological constant term $\ti{\mu}\int_Q \s{g^{(2)}}$. Here $g^{(2)}$ represents the metric in the 2d gravity. Note that in this case, the equation of motion of two dimensional metric leads to 
\ba
T^{CFT}_{ij}+\ti{\mu}g^{(2)}_{ij}=0
\ea
In particular this guarantees 
\ba
T^{CFT}_{tx}=0,  \label{reflecg}
\ea
at the interface (i.e. $\de\Sigma=\de Q$) where the 2d gravity is coupled to the 2d CFT. 
This reproduces the correct boundary condition of energy stress tensor in the BCFT, which means the complete reflection of energy flux at the boundary.
\footnote{The reflection of null geodesics in the setup of AdS/BCFT was previously discussed in \cite{Reeves:2021sab}.}
We will discuss this boundary condition of the stress tensor further in the next section.

In this treatment, the total partition function in the 2d induced gravity is expressed as
\ba
Z_{Ind}=\int [Dg^{(2)}]\int [D\Phi_{CFT}] e^{-\ti{\mu}\int_{Q} \s{g^{(2)}}} e^{-S_{CFT}(g^{(2)},\Phi_{CFT})},
\ea
where $S_{CFT}$ is the CFT action and $\Phi_{CFT}$ represents matter CFT fields on $Q$. If we first integrate out the CFT fields $\Phi_{CFT}$, then, follwing the well-known fact \cite{Polyakov:1981rd}, we obtain\footnote{In a setup for $\lambda<0$, this was evaluated in the gravity dual of path-integral optimization \cite{Boruch:2020wax,Boruch:2021hqs}. 
We can find the analytical solution to the Einstein equation for any $h_{ij}$ because the bulk solution should always be locally AdS$_3$.
In our case, we get the action with minus sign compared with
eq.(18) in \cite{Boruch:2020wax}.} the (minus) Liouville action $I_L[h]$ if we take the UV limit $e^{2\phi}\gg \ep^2$: 
\ba
I_{L}[h]\simeq -\frac{c}{24\pi}\int d\tau dx  \left[(\de_\tau\phi)^2+(\de_x\phi)^2+\mu e^{2\phi}+R\phi\right],
\label{LVef}
\ea
so that we have
\ba
Z_{Ind}=\int [Dg^{(2)}]e^{-I_L[h]},
\ea
where we performed the 2d coordinate transformation such that the metric on $Q$ is given by 
\ba
ds^2=e^{2\phi}(d\tau^2+dx^2).
\ea
The potential coefficient $\mu$ comes from $\ti{\mu}$ plus quantum corrections.
This has a wrong sign of the kinetic term compared with the normal Liouville CFT and this is indeed expected as the effective theory for a 2d CFT on a curve space.  If we choose 
\ba
\mu \, = \, \frac{1}{\delta^2} \, = \, \frac{1}{(1+\lambda^2)\ep^2} \, ,
\ea
then we get the expected metric (\ref{methet}) from the Liouville equation of motion with the background solution $e^{2\p} = (\m x^2)^{-1}$. 

Indeed, we can show that in the UV limit $e^{2\phi}\gg \ep^2$, 
the 2d gravity action computed from the gravity dual, namely $I_G[h]$, agrees with the Liouville action $I_L[h]$ in (\ref{LVef}).
Since this calculation is essentially identical to earlier works \cite{Takayanagi:2018pml,Boruch:2020wax,Boruch:2021hqs}, we will not repeat it here. This argument shows that the effective 2d gravity will be well approximated by the Liouville gravity when $\lambda$ is very large.  This identification of 2d induced gravity also agrees with the observation in \cite{Akal:2021foz} where the energy flux in the moving mirror model was found to be explained by the Liouville gravity.
For finite $\lambda$, there are higher derivative corrections non-linearly as in eq.(18) in \cite{Boruch:2020wax}. As pointed out in \cite{Omiya:2021olc}, we also expect additional non-local interactions which are expected to enhance when $\lambda$ is not large.

We can covariantize the Liouville action using the Polyakov action \cite{Polyakov:1981rd}:
\ba
I_G[h]=\frac{c}{96\pi}\int d\tau dx \s{h} \left[R\frac{1}{\Box}R-4\mu\right],
\ea
where we expressed the 2d scalar Laplace operator as
$\Box$. By introducing an auxiliary field $\vp$, we can rewrite this in a local form:
\ba
I_G[h,\vp]=-\frac{c}{24\pi}\int d\tau dx \s{h}\left[\vp\Box\vp+R\vp+\mu \right].
\label{LVeff}
\ea
This is an example of dilaton gravity in 2d, where there is a kinetic term for the dilaton as opposed to the JT gravity. As opposed to the Liouville theory, here both the scalar $\vp$ and metric $h$ are dynamical. This provides a covariant induced gravity action which is dual to the boundary dynamics of a 2d BCFT via the Island/BCFT correspondence.

\section{Energy Flux and Replica Wormholes}
\label{sec:replica wormholes}
In this section, we would like to give a justification of the extremization equation~(\ref{mirror}) we found in the island prescription
by studying the replica wormholes \cite{Almheiri:2019qdq} in this two-dimensional system. 
We also explicitly show that the boundary condition of energy stress tensor in a BCFT matches with that derived from a 2d CFT coupled to the 2d induced gravity.
In this section, we denote $a=-s$ and $b=l$ just to follow the notation of \cite{Almheiri:2019qdq}.
Therefore, the bath CFT subregion $A$ is defined by $b<x <\inf$ and the island region is defined by $-\inf < x < -a$.

Before discussing the replica wormhole, it's useful to introduce several coordinate changes.
First we move to the light cone coordinates
	\begin{align}
		x^+ \, = \, i \t + x \, , \qquad x^- \, = \, - i \t + x \, .
	\end{align}
In this coordinates, we have the metric
	\begin{align}
		ds_2^2 \, = \, e^{2\phi(x^+, x^-)} dx^+ dx^- \, ,
	\end{align}
with
	\begin{align}
		e^{2\phi(x^+,x^-)} \, &= \, \frac{4\d^2}{(x^++x^-)^2} \qquad {\rm for} \quad x^++x^-< -2\d \, , \\
		e^{2\phi(x^+,x^-)} \, &= \, 1 \qquad \qquad \qquad \ \, {\rm for} \quad x^++x^-\ge -2\d \, .
	\end{align}
Next, we would like to bring the boundary between the gravitating region and the bath CFT to a periodic circle.
For the gravitating region, this can be implemented by the conformal transformation 
\footnote{Obviously the $y$ coordinate introduced here and we use for the entire this section is differ from the $y$ coordinate used in the previous sections.}
	\begin{align}
		x^+ \, = \, \tanh\left( \frac{\pi y}{\b} \right) \, , \qquad x^- \, = \,\tanh\left( \frac{\pi \bar{y}}{\b} \right) \, ,
	\end{align}
which brings the metric in the following form
	\begin{align}
		ds_{\rm in}^2 \, = \, \left( \frac{2\pi \d}{\b} \right) \frac{dy d\bar{y}}{\sinh^2\left( \frac{\pi(y+\bar{y})}{\b} \right)} \, \qquad {\rm for} \quad y+\bar{y}< -2\d \, .
	\end{align}
For the bath CFT region, we cannot use this conformal transformation and this fact is related to the conformal welding problem discussed in \cite{Almheiri:2019qdq}.
Therefore, for the  bath CFT region, we simply set $y=x^+$ and $\bar{y}=x^-$, which gives
	\begin{align}
		ds_{\rm out}^2 \, = \, dy d\bar{y} \, \qquad {\rm for} \quad y+\bar{y}\ge -2\d \, .
	\end{align}
Furthermore, we also use the following coordinates
	\begin{align}
		w \, = \, e^{\frac{2\pi y}{\b}} \qquad ({\rm for\ \, in}) \, , \qquad v \, = \, e^{\frac{2\pi y}{\b}} \qquad ({\rm for\ \, out})
	\end{align}
which gives
	\begin{align}
		ds_{\rm in}^2 \, &= \ \frac{4\d^2 dw d\bar{w}}{(1-|w|^2)^2} \, , \\
		ds_{\rm out}^2 \, &= \, \left( \frac{\b}{2\pi} \right)^2 \frac{dv d\bar{v}}{|v|^2} \, .
	\end{align}

The crucial ingredient for the discussion of replica wormhole is the energy flux equation at the intersection between the gravitational region and the bath CFT.
Let us first review this energy flux equation for the case of JT gravity coupled to conformal matter fields \cite{Maldacena:2016upp, Engelsoy:2016xyb, Almheiri:2019psf}.
This system is defined by the total action $I_{\rm tot} = I_0+I_{\rm JT}+I_{\rm M}$ with
	\begin{align}
		I_0 \, &= \, - \frac{\p_0}{16\pi G} \left[ \int d^2x \sqrt{g} R \, + \, 2 \int dx \sqrt{h} K \right] \, \\
		I_{\rm JT} \, &= \, - \frac{\p_0}{16\pi G} \left[ \int d^2x \sqrt{g} \p (R+2) \, + \, 2 \int dx \sqrt{h} \p K \right] \, \\
		I_{\rm M} \, &= \, I_{\rm CFT}[g, \c] \, ,
	\end{align}
where $\c$ denotes a set of the matter fields.	
$I_0$ is just a topological contribution and we consider the self-interacting matter fields; namely $I_{\rm CFT}$ does not contain the dilaton field $\p$.
Therefore, the variation of the dilaton gives $R=-2$, so that the background is AdS$_2$ and we can take the Poincare coordinates
	\begin{align}
		ds_{\rm AdS_2}^2 \, = \, \frac{d\t^2 + d\h^2}{\h^2} \, .
	\end{align}
On the other hand, the variation with respect to the metric gives the dilaton equation 
	\begin{align}
		\nabla_\m \nabla_\n \p \, - \, g_{\m\n} \nabla_\sigma \nabla^\sigma \p \, + \, g_{\m\n} \p \, = \, - 8\pi G T^{(M)}_{\m\n} \, ,
	\label{dilaton-eq}
	\end{align}
where $T^{(M)}_{\m\n}$ is the matter field stress tensor.
In order to consider the dynamical boundary \cite{Maldacena:2016upp}, we parametrize the boundary by the boundary time $u$ as ($\t(u), \h(u)$).
The boundary condition of the metric gives us
	\begin{align}
		\frac{1}{\e^2} \, = \, \frac{\t'^2+\h'^2}{\h^2} \quad \Rightarrow \quad \h \, = \, \e \t' \, + \, \mathcal{O}(\e^3) \, ,
	\end{align}
where the prine denotes a derivative with respect to $u$ and the boundary condition of the dilaton gives
	\begin{align}
		\p \big|_{\rm bdy} \, = \, \frac{\bar{\p}_r}{\e} \, .
	\end{align}
The energy flux equation comes from the ($\t, \h$) component of the dilaton equation (\ref{dilaton-eq}), which is explicitly written as
	\begin{align}
		\pa_\t \pa_\h \p  \, = \, - 8\pi G T^{(M)}_{\t \h} \, .
	\end{align}
By using $\pa_\t = (1/\t') \pa_u$ and introducing a normal derivative of the boundary by $\pa_n=\e (\h' \pa_\t - \t' \pa_\h)$, we can rewrite the energy flux equation as 
	\begin{align}
		- \pa_u \bigg( \frac{\sqrt{h}}{8\pi G}  \, \pa_n \p \bigg)  \, = \, - (\t')^2 \, T^{(M)}_{\t \h} \, ,
	\end{align}
up to a singular contribution.
For JT gravity, we note that the ADM energy is given by \cite{Maldacena:2016upp}
	\begin{align}
		M \, = \, - \, \frac{\sqrt{h}}{8\pi G} \big(\pa_n \p \, + \, \cdots \big)\big|_{\rm bdy} \, = \, \frac{\bar{\p}_r}{8\pi G} \, \{\t, u \} \, ,
	\end{align}
where the ellipsis denotes a singular contribution.
(In appendix~\ref{app:adm energy}, we gives a short summary for a derivation of the ADM energy in JT gravity. For more complete discussion, see \cite{Maldacena:2016upp}.)
Therefore, the flux equation is now expressed in terms of the change of the ADM energy as \cite{Maldacena:2016upp, Engelsoy:2016xyb, Almheiri:2019psf}
	\begin{align}
		\pa_u M  \, = \, - (\t')^2 \, T^{(M)}_{\t \h} \, = \, - i (\t')^2 \big( T_{x^+x^+} - T_{x^-x^-} \big) \, .
	\label{JT_energy_flux}
	\end{align}
The discussion of replica wormhole in this system (i.e. JT gravity coupled to conformal matter fields) is summarized in appendix~\ref{app:replica wormholes}.

Now we come back to our AdS/BCFT set-up.
The induced gravity on the brane actually does not have the Schwarzian boundary action as we discussed in section~\ref{sec:2d gravity}.
Therefore, in the present set-up, the boundary energy flux equation is simply given by the bath CFT energy-momentum tensor as
	\begin{align}
		T_{yy}(i u) \, - \, T_{\bar{y}\bar{y}}(-i u) \, = \, 0 \, ,
	\label{energy flux eq}
	\end{align}
where $2i u =y - \bar{y}$.
Again this energy flux equation perfectly agrees with the energy reflection equation in a BCFT point of view.

Under the conformal transformation
	\begin{align}
		y \, \to \, z \, \equiv \, F(v) \, ,
	\end{align}
the energy-momentum tensor transforms as
	\begin{align}
		T_{yy}(iu) \, \to \, e^{\frac{4\pi y}{\b}} \left[ \left( \frac{dF(e^{\frac{2\pi y}{\b}})}{dv} \right)^2 T_{zz} \, - \, \frac{c}{24\pi} \left\{ F(e^{\frac{2\pi y}{\b}}), v \right\} \right] \, .
	\end{align}
Now we consider the replicated geometry. For this geometry, the uniformizing map is given by $z \to \tilde{z} \equiv z^{1/n}$, such that $T_{\tilde{z} \tilde{z}}=0$.
Therefore, the energy-momentum tensor on the $z$-plane is
	\begin{align}
		T_{zz} \, = \, - \frac{c}{24\pi} \left\{ z^{1/n}, z \right\} \, = \, - \frac{c}{48\pi} \left( 1- \frac{1}{n^2} \right) \frac{1}{z^2} \, .
	\end{align}
Combining all, now the energy flux equation is written as
	\begin{align}
		0 \, = \, i e^{\frac{4\pi i u}{\b}} \left[ \frac{(n^2-1)}{2n^2} \frac{F'(e^{\frac{2\pi i u}{\b}})^2}{F(e^{\frac{2\pi i u}{\b}})^2} \, + \, \left\{ F(e^{\frac{2\pi u}{\b}}), v \right\} \right] \, + \, c.c. 
	\label{EF-eq}
	\end{align}

In order to study the $n \to 1$ limit, it's convenient to set
	\begin{align}
		F(v) \, = \, \frac{v-A}{B-v} \, , \qquad A \, = \, e^{-\frac{2\pi a}{\b}} \, , \quad B \, = \, e^{\frac{2\pi b}{\b}} \, .
	\label{F(v)}
	\end{align}
We note that for this choice of $F(v)$, we have
	\begin{align}
		\left\{ F(v), v \right\} \, = \, 0 \, .
	\end{align}
This is simply because the choice of $F(v)$ in (\ref{F(v)}) is a special case of M\"{o}bius transformation
and for any M\"{o}bius transformation the Schwarzian derivative is zero.
Therefore, the energy flux equation becomes
	\begin{align}
		\mathcal{F} \, = \, 0 \, ,
	\end{align}
where
	\begin{align}
		\mathcal{F} \, &= \, i e^{\frac{4\pi i u}{\b}} \frac{F'(v)^2}{F(v)^2} \, + \, c.c. \nn\\
		&= \, i \frac{e^{\frac{4\pi i u}{\b}}(A-B)^2}{(e^{-\frac{2\pi i u}{\b}} - A)^2(e^{\frac{2\pi i u}{\b}} - B)^2} \, + \, c.c. 
	\label{mathcalF}
	\end{align}
Furthermore, Fourier transforming the energy flux equation from $u$ to $k$, the $k=1$ equation reads
	\begin{align}
		0 \, = \, \int_0^{\b} du \, e^{-\frac{2\pi i u}{\b}} \, \mathcal{F} \, ,
	\end{align}
Performing the $\t$-integration, this is written as
	\begin{align}
		0 \, = \, \frac{\sinh(\frac{\pi(a-b)}{\b})}{\sinh(\frac{\pi(a+b)}{\b})} \, .
	\end{align}
In order to compare with the quantum extremal surface condition (\ref{mirror}), we need to go back to the infinite straight line boundary.
This is simply obtained by taking $\b \to \inf$ and this leads to
	\begin{align}
		0 \, = \, \frac{a-b}{a+b} \, .
	\end{align}
This agrees with the quantum extremal surface condition (\ref{mirror}) and set the location $a$ by $a=b$.

In this section, we have lengthily discussed replica wormholes mainly following \cite{Almheiri:2019qdq} to obtain the externalization equation $a=b$.
However, we can summarize what we have done in this section much shortly without replica wormholes.
The crucial equation is again the energy flux conservation equation (\ref{energy flux eq}).
More precisely we require this energy flux conservation on the $z$ plane:
	\begin{align}
	    T_{zz}(z) \, - \, T_{\bar{z}\bar{z}}(\bar{z}) \, = \, 0 \, .
	\end{align}
By transforming 
	\begin{align}
	    z \, \to \, \tilde{z} \, = \, \left( \frac{x^+ - a}{b -x^+} \right)^{\frac{1}{n}} \, , \qquad 
	    \bar{z} \, \to \, \bar{\tilde{z}} \, = \, \left( \frac{x^- - a}{b -x^-} \right)^{\frac{1}{n}} \, ,
	\end{align}
we consider the vacuum state on the $\tilde{z}$ plane i.e. $ T_{\tilde{z}\tilde{z}}(\tilde{z}) = T_{\bar{\tilde{z}}\bar{\tilde{z}}}(\bar{\tilde{z}}) = 0$.
Therefore, the stress tensor on the $z$ plane is simply given by the Schwarzian derivative as
	\begin{align}
	    T_{zz}(z) \, &= \, - \frac{c}{24\pi} \big\{ \tilde{z} , z \big\}
	    \, = \, - \frac{c(1-\frac{1}{n^2})}{48\pi} \frac{(a-b)^2}{(a-x^+)^2(b-x^+)^2} \, , \\
	    T_{\bar{z}\bar{z}}(\bar{z}) \, &= \, - \frac{c}{24\pi} \big\{ \bar{\tilde{z}} , \bar{z} \big\}
	    \, = \, - \frac{c(1-\frac{1}{n^2})}{48\pi} \frac{(a-b)^2}{(a-x^-)^2(b-x^-)^2} \, ,
	\end{align}
where $x^{\pm} = \pm i \t - \d$. Since we require $ T_{zz}(z) = T_{\bar{z}\bar{z}}(\bar{z}) $ for any value of $\t$,
this implies we have to have $a=b$. This is the externalization equation.

\section{AdS/BCFT and One Point Function}
\label{sec:one point function}

In general, one point functions in BCFT are non vanishing  \cite{Cardy:1984bb}. For a scalar primary operator, the one point function looks like
\ba
\la O(y)\lb=\frac{{\cal N}}{|y|^{\Delta}},  \label{onepp}
\ea
where $|y|$ is the distance from the boundary and $\Delta$ is the conformal dimension of the scalar primary operator $O$. We also write the overall normalization as ${\cal N}$. Below we will see that 
in order to reproduce this non-vanishing one-point function, we need to consider a bulk gravity solution with a non-trivial expectation value of a bulk scalar (see also earlier work \cite{Fujita:2011fp,Kastikainen:2021ybu}). Moreover, we need to explain the non-vanishing one point function from the 2d gravity picture to justify the Island/BCFT correspondence. Again we focus on two dimensional BCFTs and we employ the Euclidean signature. We will first study $\Delta=2$ case where we can obtain analytical results and later examine $\Delta\neq 2$ case numerically. The basic guide line in the gravity dual is that the 3d metric is foliated by AdS$_2$, which explains the boundary conformal invariance. 

\subsection{Massless Bulk Scalar $\Delta=2$}

As a special case where we have an analytical solution, let us start with
the scalar operator with the dimension $\Delta=2$, which is dual to
a massless bulk scalar $\phi$ in the AdS/BCFT setup.  This scalar field is described by the standard action:
\ba
I=-\frac{1}{16\pi G_N}\int d^3x \s{g}(R-g^{ab}\de_a\phi\de_b\phi+2).
\ea
As shown in \cite{Bak:2007jm}, this has the Janus solution 
\ba
&& ds^2 \, = \, d\r^2+ f(\r) \,ds^2_{\rm AdS_2} \, , \qquad f(\r) \, = \, \frac{1+\s{1-2\gamma^2}\cosh 2\r}{2} \, , \no
&& \phi \, = \, \phi_0+\frac{1}{\s{2}}\log\left[\frac{1+\s{1-2\gamma^2}+\s{2}\gamma\tanh \r}{1+\s{1-2\gamma^2}-\s{2}\gamma\tanh \r}\right],  \label{janusol}
\ea
where $ds^2_{AdS_2}=\frac{d\eta^2+d\tau^2}{\eta^2}$
is the Euclidean AdS$_2$ metric. The parameter $\gamma$ describes the amount of the Janus deformation 
such that we have the pure AdS$_3$ solution at $\gamma=0$.

We will show that we can obtain a class of setups of AdS$_3$/BCFT$_2$ from this solution. 
We assume the EOW brane $Q$ is at $\r=\r_*$, such that the gravity dual extends in the region $\r_*<\r<\infty$. Clearly, this background has the isometry of AdS$_2$ which is dual to the boundary conformal invariance.

For the scalar field $\phi$, we assume the linear interaction on the brane
\ba
I_{bdy}=\frac{a}{8\pi G_N}\int_{Q}\s{h} \phi,
\ea
where $a$ is a coupling constant and $h_{ij}$ is the induced metric on $Q$.
Combined with the original action, under the variation of $\phi$, the total action $I+I_{bdy}$ leads to the Neumann-like boundary condition at $Q$:
\ba
\de_\r\phi|_{\r=\r_*}=a. \label{EOWbca}
\ea
This is satisfied if we set the parameter $\gamma$ is related to $a$ via
\ba
\frac{2\gamma(1+\s{1-2\gamma^2})}{(1+\s{1-2\gamma^2})^2\
\cosh^2 \r_*-2\gamma^2\sinh^2 \r_*}=a.
\ea
The Neumann boundary condition of the gravity  coupled to the scalar field reads
\ba
K_{ij}-h_{ij}K=-(T+a\phi)h_{ij}.\label{EOWbcaa}
\ea
This is satisfied if 
\ba
T+a\phi(\r_*) \, = \, - \frac{f'(\r_*)}{2f(\r_*)}. \label{EOWbcb}
\ea
We find for $\r_*\to -\infty$ we have $T\to \infty$ and vice versa. For a given value of $T$ and $a$ we can determine the values of $\rho_*$ and $\gamma$, which give the AdS/BCFT setup. For example, the boundary entropy in this model can be found from the holographic entanglement entropy as we did in (\ref{EEISA})  and thus it is given by 
\ba
S_{bdy}=\frac{c}{6}\rho_*.
\ea

Now let us calculate the one-point function. In the AdS/CFT on the Poincare coordinate,
\ba
ds^2=\frac{d\t^2+dy^2+dz^2}{z^2},
\ea
the bulk scalar behaves near the AdS boundary $z\to 0$:
\ba
\phi(\t,y,z)\simeq J(\t,y) z^{2-\Delta} + \ap(\t,y) z^\Delta+\ddd.
\ea
Here $J$ is interpreted as the source such that it adds the linear interaction $\int d\t dy J(\t,y)O(\t,y)$ and $\ap$ corresponds to the expectation value as 
\ba
\ap(\t,y)=\la O(\t,y)\lb.
\ea
In our solution (\ref{janusol}) we find  in the $\r\to\infty$ limit:
\ba
\phi\simeq \phi_*-\frac{4\gamma(1+\s{1-2\gamma^2})}{(1+\s{1-2\gamma^2})^2-2\gamma^2}e^{-2\r}.
\ea

Since the Poincare coordinate is related to the AdS$_2$ slice coordinate via $\frac{y}{z}\simeq \sinh \r$, we find 
\ba
\la O(\t,y)\lb \, = \, - \frac{\gamma(1+\s{1-2\gamma^2})}{(1+\s{1-2\gamma^2})^2-2\gamma^2}\cdot \frac{1}{y^2}.
\ea
Since the massless field corresponds to $\Delta=2$, this result indeed agrees with the general form (\ref{onepp}). Note that the coefficient of one point function goes to $\pm \infty$ when $\gamma\to\mp 1/\s{2}$. Therefore, we can take any real values of the boundary entropy $S_{bdy}$ and the one-point function coefficient ${\cal N}$ in this holographic model.

\begin{figure}[t!]
    \begin{center}
        \ \includegraphics[scale=0.5]{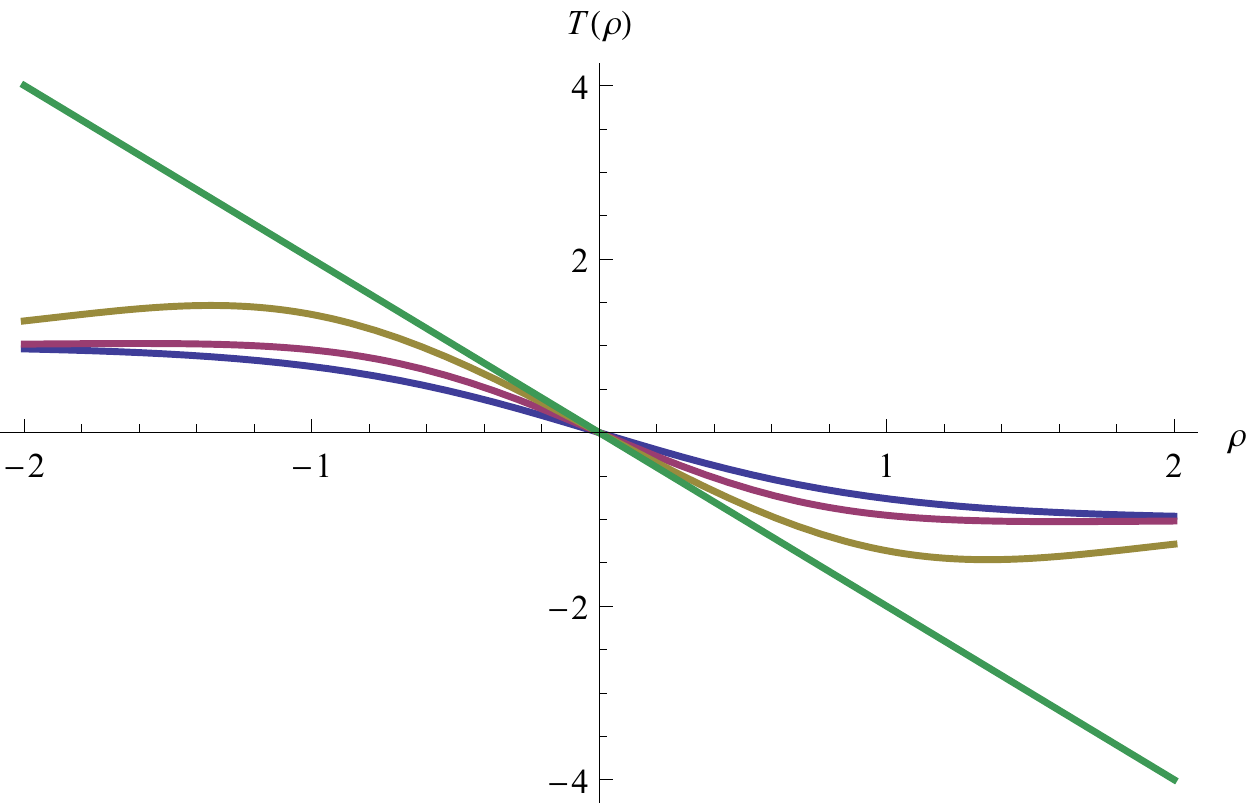} \ \ \includegraphics[scale=0.5]{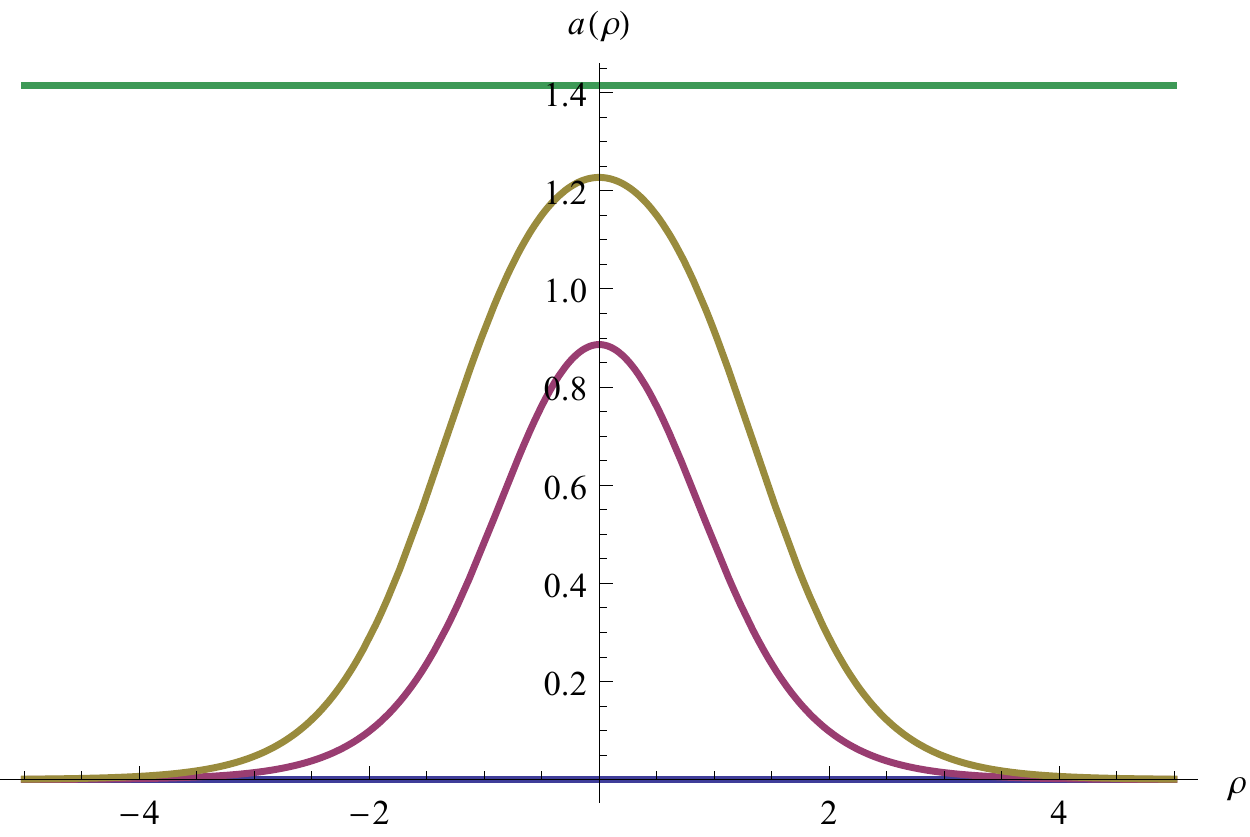} 
    \end{center}
    \caption{The tension parameter $T$ (left) and the boundary condition parameter $a$ (right) as functions of $\rho$ for fixed values of $\gamma$.  The blue, red, yellow and green curves correspond to 
    $\gamma=0$, $\gamma=0.9/\s{2}$, $\gamma=0.99/\s{2}$ and $\gamma=1/\s{2}$, respectively. We set $\phi_0=0$.    }   \label{fig:parameterjanus}
\end{figure}

On the other hand, the allowed range of the parameters $(T,a)$ in this model 
is non-trivially limited as plotted their behaviors  as functions $\rho$ for various values of $\gamma$ in Figure~\ref{fig:parameterjanus}. For example, $a$ is always bounded as $|a|\leq \s{2}$, where the equality is saturated in the extreme limit $\gamma\to \pm 1/\s{2}$.
It is also interesting to note that in the original AdS/BCFT model \cite{Takayanagi:2011zk,Fujita:2011fp} without any matter fields, the range of the tension $T$ is limited  to $|T|\leq 1$. However, in our generalized model with a scalar, this is no longer true.
In the extreme limit $\gamma\to 1/\s{2}$, we find $T\to -2\rho_*$, which is no longer bounded.

\subsection{Massive Bulk Scalar: $\Delta\neq 2$}

We would like to extend the previous analysis to a massive scalar so that we can treat operators with $\Delta\neq 2$.
We start with the following general action:
\ba
I=-\frac{1}{16\pi G_N}\int d^3x\s{g}\left[ R-g^{ab}\de_a\phi\de_b\phi-U(\phi)\right].
\ea

The scalar equation of motion and Einstein equation read
\ba
&& -2\de_a\Big(\s{g}g^{ab}\de_b\phi\Big)+\s{g}U'(\phi) \, = \, 0 \, , \no
&& R_{ab}-\frac{1}{2}g_{ab}R+\frac{1}{2}g_{ab}U(\phi) \, = \, -\frac{1}{2}g_{ab}(\pa \p)^2+\de_a\phi\de_b\phi.
\ea
where $(\pa \p)^2 = g^{ab}\de_a\phi\de_b\phi$.

\subsubsection{Ansatz and Boundary Condition}

We impose the metric and scalar field ansatz:
\ba
&& ds^2=d\r^2+f(\r)\left(\frac{d\eta^2+d\tau^2}{\eta^2}\right),\no
&& \phi=\phi(\r).
\ea
Then the equation of motions take the following form
\ba
&&-2\de_\r(f\de_\r\phi)+fU'(\phi)=0,\label{eomsc} \\
&& (\de_\r f)^2 + 4f + 2 f^2 U(\phi) - 2 f^2 (\de_\r\phi)^2=0, \label{eina} \\
&& 2f \de_\r^2f - (\de_\r f)^2 + 2f^2 U(\phi) + 2f^2(\de_\r\phi)^2=0.
\label{einb}
\ea
The previous massless case can be obtained by choosing $U(\phi)=-2$. For a free massive scalar we have $U(\phi)=-2+m^2\phi^2$. The null energy condition reads $N^aN^b R_{ab}\geq 0$.
By choosing $N^\r=1,\ N^\eta=0,\ N^\tau=i \h /\sqrt{f}$,  this leads to 
\ba
2f + (\de_\r f)^2 - f\de_\r^2f \, \geq \, 0.  \label{nec}
\ea

We can show that the scalar field equation of motion (\ref{eomsc}) is automatically satisfied if the Einstein equations (\ref{eina}) and (\ref{einb}) hold. 
Therefore the independent equations of motion, which we need to solve, are summarized as 
\ba
&&  (\de_\r\phi)^2 \, = \, \frac{2f + (\de_\r f)^2 - f\de_\r^2f}{2f^2},\label{emma}\\
&&  \de_\r^2f + 2 + 2 f U(\phi) \, = \, 0. \label{emmb}
\ea
Note that the null energy condition (\ref{nec}) simply says the obvious fact that the right hand side of (\ref{emma}) is non-negative. For any given function $\phi(\r)$ we can find $f(\r)$ by solving (\ref{emma}) and also find the potential $U(\phi)$ by solving (\ref{emmb}). 

We would like to find a solution which satisfies the boundary condition at the AdS boundary $\rho\to \infty$:
\ba
\phi(\r)\simeq \phi_1+\ap e^{-\Delta \r}, \ \ \ f(\r)\simeq f_1 e^{2\r},  \label{bcmassives}
\ea
where we can set $\phi_1=0$ by shifting the scalar field. We also impose the boundary condition on the end of the world-brane $\rho=\rho_*$ (\ref{EOWbca}) and (\ref{EOWbcb}). The boundary behavior of $\phi(\rho)$ in (\ref{bcmassives}) shows that the one-point function in BCFT agrees with the general form (\ref{onepp})  and the coefficient  ${\cal N}$ is proportional to the parameter $\ap$.

\subsubsection{Analytical model at $\Delta=1$}

For $\Delta=1$, we can find a simple analytical model.
We choose
\ba
f(\r) \, = \, \big( A\cosh\lambda \r \big)^2 \, .
\ea
This determined the scalar field as
\ba
\de_\r\phi=\frac{\s{1-A^2\lambda^2}}{A\cosh\lambda \r}.
\ea
Thus we require $A\lambda\leq 1$.
This is solved as follows
\ba
\phi=\frac{2\s{1-A^2\lambda^2}}{A\lambda}\arctan\left[\tanh\frac{\lambda \r}{2}\right].
\ea
In the limit $\r\to\pm\infty$, it behaves as follows
\ba
\phi(\r)\simeq \frac{2\s{1-A^2\lambda^2}}{A\lambda}\left(\pm \frac{\pi}{4}\mp e^{\mp \lambda \r}\right).\label{scabeh}
\ea
Therefore this solution interpolates two criticial points  $\phi=\phi_{\pm}\equiv \pm\frac{\pi\s{1-A^2\lambda^2}}{2A\lambda}$.
The potential is found as 
\ba
U(\phi)=-2\lambda^2-\frac{1-A^2\lambda^2}{A^2}\cos^2\left[\frac{A\lambda}{\s{1-A^2\lambda^2}}\phi\right].
\ea
Note that we get the pure AdS$_3$ solution if we set $A\lambda=1$.

Around $\phi=\phi_\pm$, it is expanded as 
\ba
U(\phi)\simeq -2\lambda^2-\lambda^2(\phi-\phi_{\pm})^2.
\ea
Since the AdS radius is $L_{AdS}=\frac{1}{\lambda}$ in the limit $|\r|\to\infty$, 
we can obtain the conformal dimension of primary dual to $\phi$ as $\Delta=1$. This also agrees with the bahavior in (\ref{scabeh}).
In this model, we can limit the space to $\r_*<\r<\infty$ to get the gravity dual of the BCFT. 

\subsection{Numerical solution for free massive scalar}
We consider the free massive dilaton case by setting $U(\phi)=-2+m^2\phi^2$.
The Einstein equation and the Klein-Gordon equation are given by
	\begin{align}
		f'' - 4 f + 2 \, &= \, - 2 m^2 f \p^2 \, , \label{Ein-eq1} \\
		(f')^2 - 2f f'' + 4f^2 \, &= \, 2f^2( m^2 \p^2 + (\p')^2) \, , \label{Ein-eq2} \\
		f \p'' + f' \p' \, &= \, m^2 f \p \, . \label{KG-eq}
	\end{align}
The two equations (\ref{Ein-eq1}) and (\ref{Ein-eq2}) from the Einstein equation are not independent and can be reduced into one equation
	\begin{align}
		(f')^2 - 4 f^2 + 4f \, = \, 2f^2 \big((\p')^2 - m^2 \p^2 \big) \, . \label{Ein-eq3}
	\end{align}
Therefore, we need to solve the Klein-Gordon equation (\ref{KG-eq}) and this equation (\ref{Ein-eq3}) simultaneously for $f$ and $\p$.
For the boundary conditions (\ref{bcmassives}) for $f$ and $\p$, we impose in the $\r \to \inf$ limit
	\begin{align}
		f(\r) \, &\to \, \frac{1}{4} \, e^{2\r} \, + \, \frac{1}{2} \, - \, \frac{\a^2}{8} \, e^{2(1-\D) \r} \, + \, \cdots \, , \label{frhoa} \\
		\p(\r) \, &\to \, \a \, e^{-\D \r} \, + \, \cdots \, , 
	\end{align}
where we kept two different subleading terms for $f(\r)$ because the actual subleading term depends on whether
$\Delta<1$ ($e^{2(1-\Delta)\r}$ is dominant) or $\Delta>1$ ($1/2$ is dominant).
The coefficients of these subleading terms are fixed by consistency of the Einstein equation (\ref{Ein-eq3}).
Also the coefficient of the leading term in $f(\r)$ is adjustable by shifting $\r$ coordinate, so we choose it for later convenience. From the behavior of $\phi(\rho)$ in (\ref{frhoa}) we can confirm that the one-point function in BCFT (\ref{onepp}) can be reproduced and the coefficient  ${\cal N}$ is proportional to the parameter $\ap$.

\begin{figure}[t!]
    \begin{center}
        \ \includegraphics[scale=0.74]{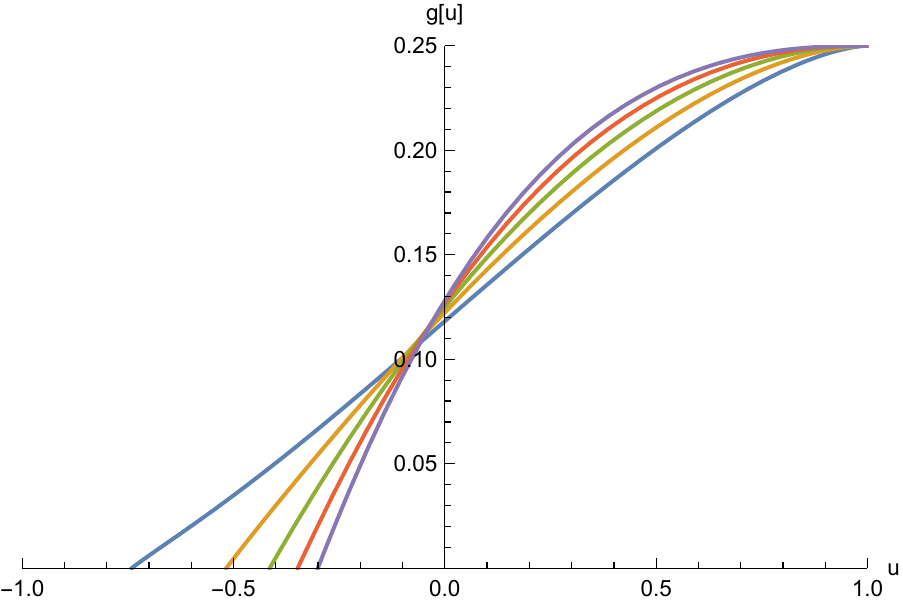} \ \ \includegraphics[scale=0.74]{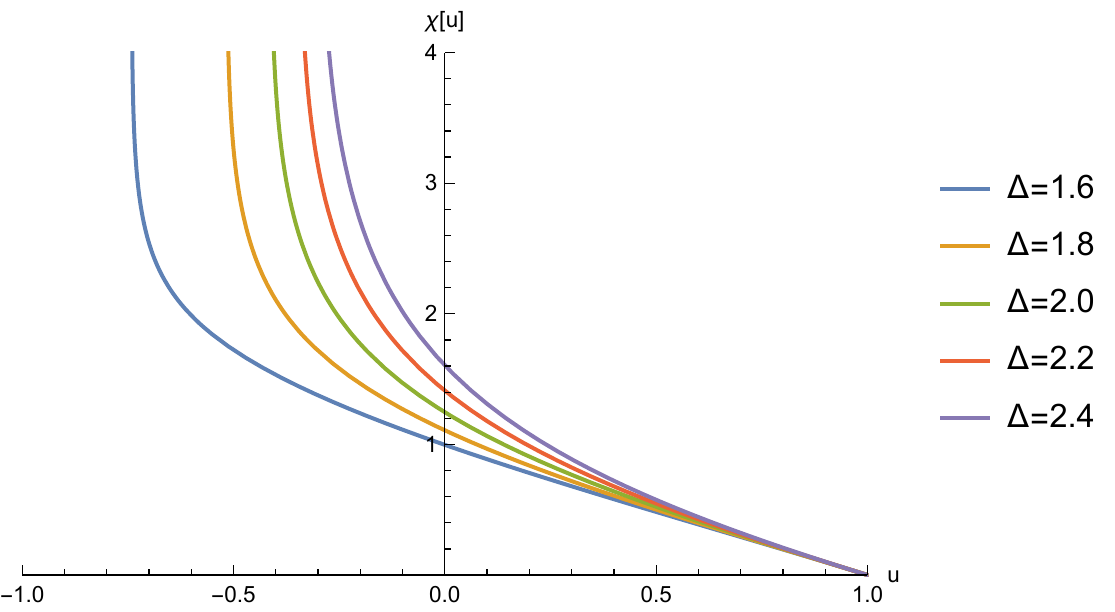} \\ \vspace*{10pt}
        \ \includegraphics[scale=0.74]{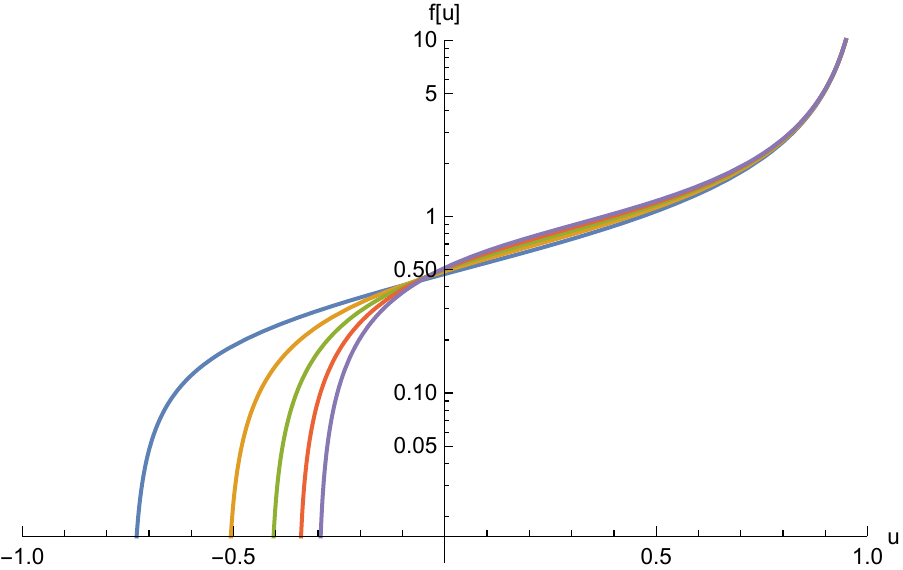} \ \ \includegraphics[scale=0.74]{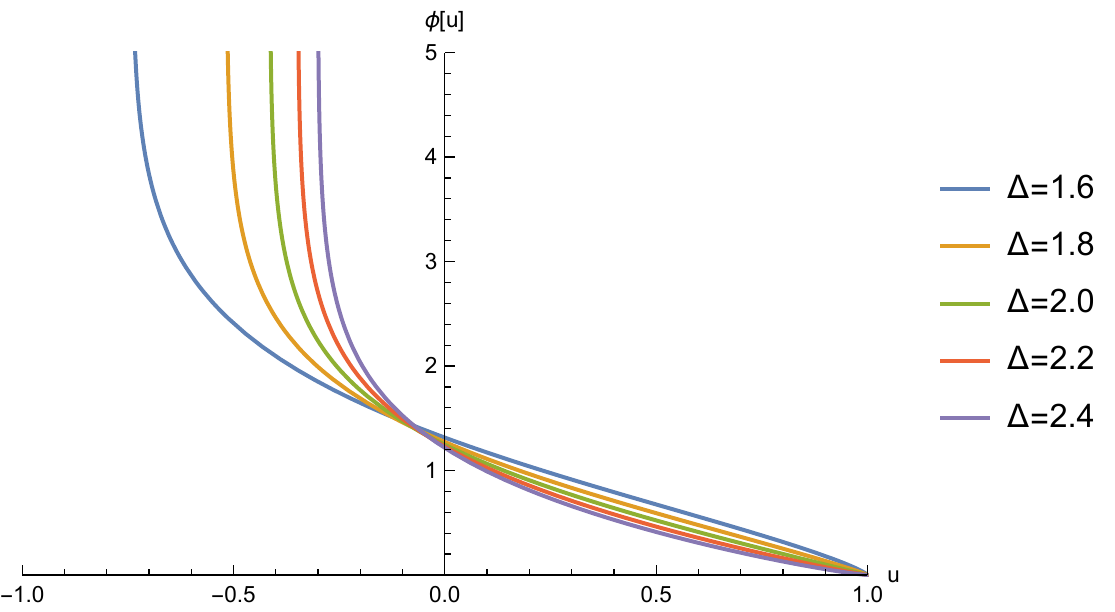} 
    \end{center}
    \caption{Numerical plots of singular solutions of $g(u)$, $\c(u)$, $f(u)$ and $\p(u)$ with $\D=\D_+$ for various dimensions in the range of $1.6 \le \D_+ \le 2.4$.
    Here we have Log plot for $f(u)$ and choose the parameter $\a = 2.0$.}
    \label{fig:singular}
\end{figure}

\begin{figure}[t!]
    \begin{center}
        \ \includegraphics[scale=0.52]{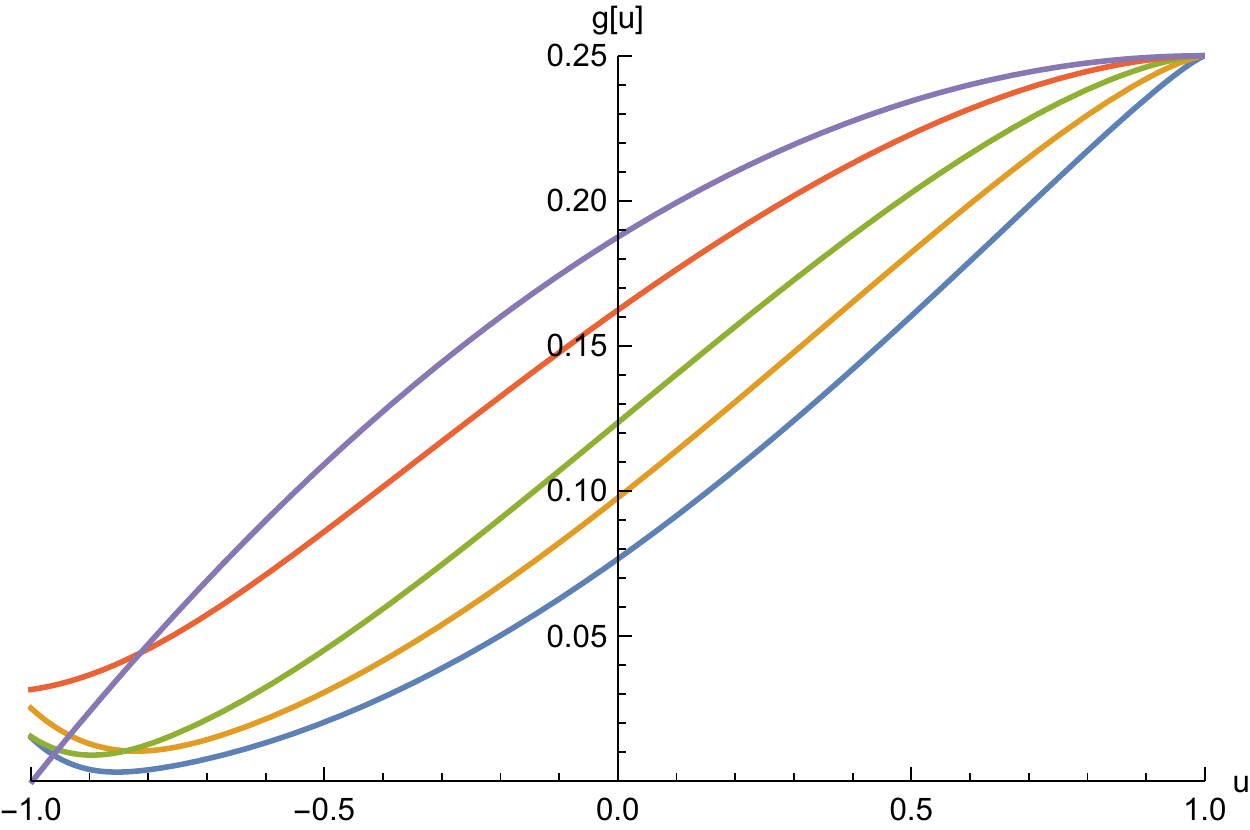} \ \ \includegraphics[scale=0.52]{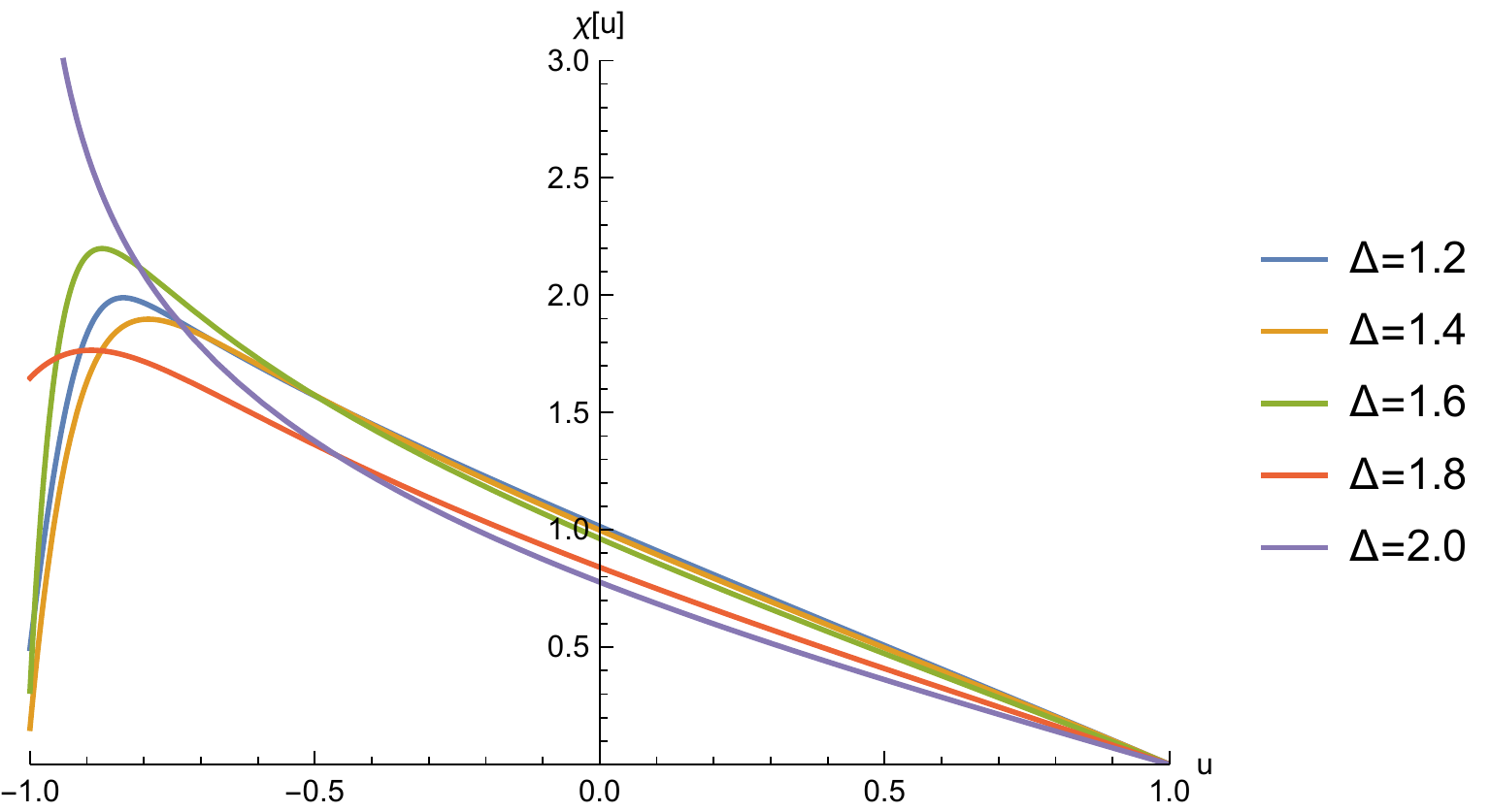} \\ \vspace*{10pt}
        \ \includegraphics[scale=0.52]{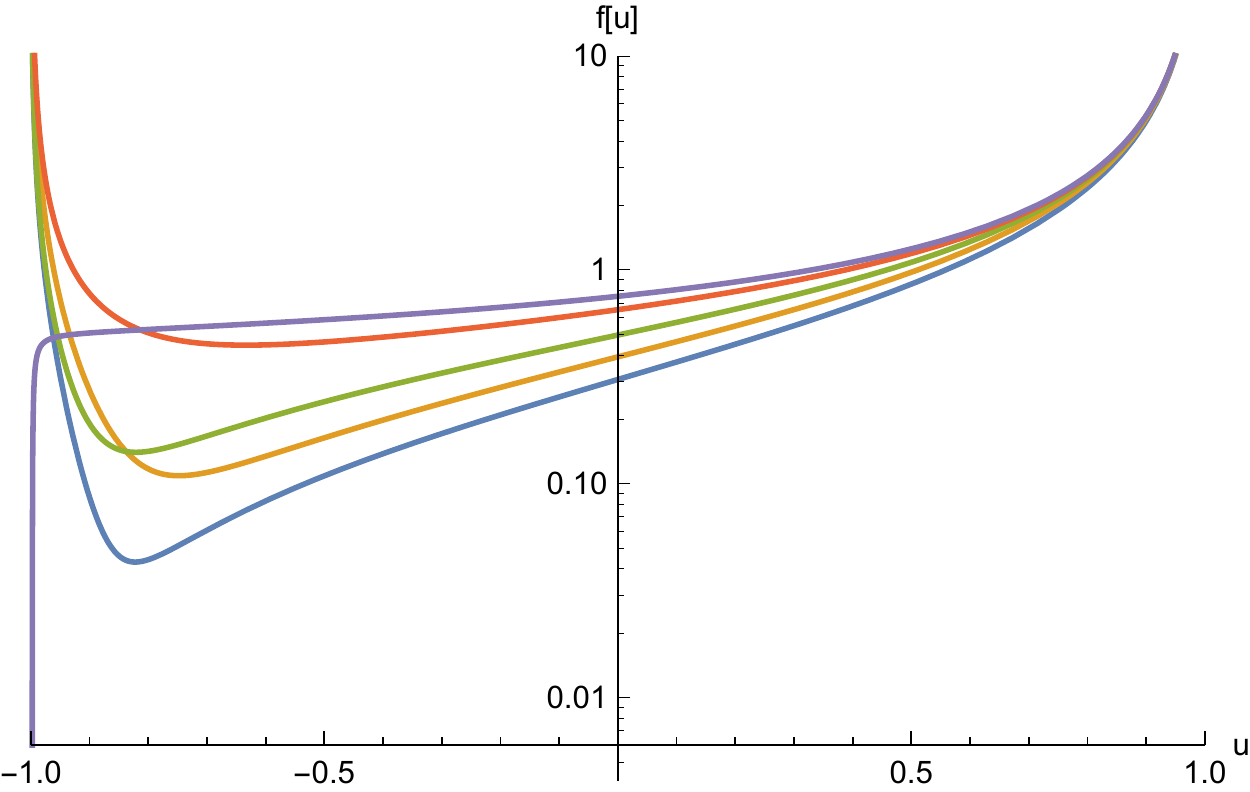} \ \ \includegraphics[scale=0.52]{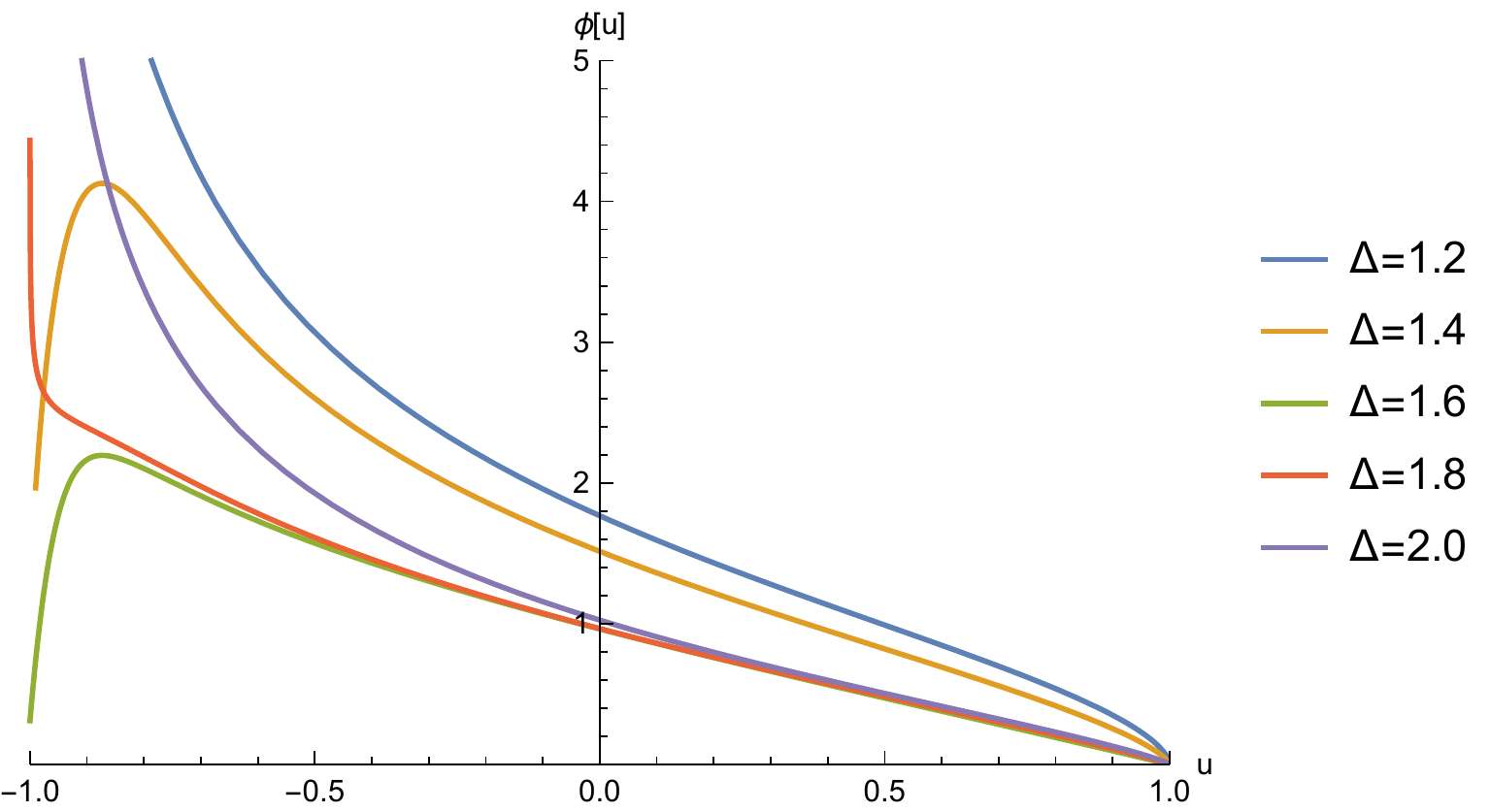} 
    \end{center}
    \caption{Numerical plots of non-singular solutions of $g(u)$, $\c(u)$, $f(u)$ and $\p(u)$ with $\D=\D_+$ for various dimensions in the range of $1.2 \le \D_+ \le 2.0$.
    At least, adjusting the values of $\a$, we can find a non-singular solution for a wide range of non-irrelevant dimension $\D$.
    Here we took $\a=2.101$ (for $\D=1.2$), $\a=2.083$ (for $\D=1.4$), $\a=1.962$ (for $\D=1.6$), $\a=1.663$ (for $\D=1.8$) and $\a=1.414$ (for $\D=2.0$).}
    \label{fig:non-singular}
\end{figure}

The numerical solutions for various small masses are shown in Figure~\ref{fig:singular} and \ref{fig:non-singular}.
In order to perform numerical evaluation, we have to map the coordinate and the field into finite space.
To this end, we introduced
	\begin{align}
		u \, = \, \tanh \r \, ,
	\end{align}
which map the coordinate into $-1 \le u \le 1$ and
\footnote{
We could use $\p(u) = \left(\frac{1-u^2}{4} \right)^{\frac{\D}{2}} \c(u)$, but it seems $\frac{\D}{2}-1$ factor is numerically more stable \cite{Jansen:2017oag}.
}
	\begin{align}
		f(u) \, = \, \frac{4 g(u)}{1-u^2} \, , \qquad \p(u) \, = \, \left(\frac{1-u^2}{4} \right)^{\frac{\D}{2}-1} \c(u) \, ,
	\end{align}
where we choose $\D= \D_{+}=1+\sqrt{1+m^2}$.
\footnote{The choice $\D= \D_{-}=1-\sqrt{1+m^2}$ corresponds to excited states in the boundary theory, which might be interesting for other topics,
but here we focus on the choice $\D= \D_{+}=1+\sqrt{1+m^2}$ which corresponds to the vacuum state in the boundary theory.}
In this paper, we focus on the masses $m^2 > -1$ above the BF bound \cite{Breitenlohner:1982bm, Breitenlohner:1982jf},
so in terms of the dimensions we consider $\D_+ >1$.
In terms of $g$ and $\c$, the boundary conditions are written as
	\begin{align}
		g(u) \, &\to \, \frac{1}{4} \, - \, \frac{\a^2}{8} \, \left(\frac{1-u^2}{4} \right)^{\D} \, + \, \cdots \, , \\
		\c(u) \, &\to \, \a \left(\frac{1-u^2}{4} \right) \, + \, \cdots \, ,
	\end{align}
for $u \to 1$. In particular this means that $\c(u) \to 0$ and $\c'(u) \to -\a/2$ in the $u \to 1$ limit.

In Figure~\ref{fig:singular}, we plotted singular solutions for various dimensions in the range of $1.6 \le \D_+ \le 2.4$.
For each value of $\D_+$ (or $m^2$) we see a naked singularity at $u=u_s$ where $g(u_s)=0$ (i.e. $f(u_s)=0$).
From (\ref{emma}), at such location we have $\p'(u_s)$ diverging, so the location is singular.
Such a naked singularity must be prohibited for the usual AdS/CFT without an EOW brane \cite{Bak:2003jk, Bak:2007jm}.
However, for our current study with an EOW brane, the existence of the naked singularity just means that we have to place the EOW brane before this singularity ($u_s < u_*$).
Then, the presence of this singularity behind the EOW brane does not cause any problem.
In general, as we decrease the value of $\a$, the singular location $u_s$ moves towards the other boundary $u \to -1$ and it eventually becomes a non-singular solution.
On the other hand, if we increase the value of $\a$, the singular location $u_s$ moves towards the first boundary $u \to +1$;
therefore the allowed bulk region becomes narrower.

In Figure~\ref{fig:non-singular}, we plotted non-singular solutions for the entire bulk region ($-1\le u \le 1$) with non-irrelevant dimensions $\D_+$.
We have not investigated the whole solution space (which is spanned by $\D_+$ and $\a$),
but at least adjusting values of $\a$ appropriately, we could find such non-singular solutions for a wide range of non-irrelevant dimensions $\D_+$.

We can read off the holographic energy stress tensor $T_{ij}$ by rewriting the above solution into the Fefferman-Graham expansion:
\ba
&& ds^2=\frac{dz^2+\g_{ij}(x,z)dx^i dx^j}{z^2},\no
&& \g_{ij}(x,z)\simeq \delta_{ij}+T_{ij}(x)z^2+O(z^3).
\ea
From this analysis, we can confirm from the behavior of $f(\r)$ given by (\ref{frhoa}) that $T_{ij}$ is vanishing.
We write $(x^1,x^2)=(\tau, y)$. Therefore the above numerical solution indeed corresponds to the vacuum state of the BCFT.

\subsection{One Point Function and Island/BCFT Correspondence}

Finally, we would like to come back to the Island/BCFT correspondence.
Since one point functions (\ref{onepp}) are generally non-vanishing in BCFTs, the same should be true in the other description where a 2d CFT is coupled to a 2d induced gravity. The latter theory is described by a 2d CFT on the union of 
AdS$_2$ and a flat half plane as in (\ref{methet}) with the metric is dynamical only on the AdS$_2$ region. One may think that it is clear that the one-point function of a primary is vanishing as is true in 2d CFTs on $R^2$. Moreover, the dynamical metric on AdS$_2$ does not seem to change the situation. 

However, if we look into the AdS/BCFT solution shown in previous subsections, the non-trivial profile of a scalar field plays a crucial role to reproduce the one-point function. Indeed, we find that the value of the scalar field on the EOW brane $Q$ given by $\phi(\rho_*)$ is non-vanishing when $\alpha\neq 0$. This means that there is a source to $O(x)$ on the 2d gravity on $Q$. We can write this source by adding $\hat{\ap}\int d\tau' dx'\s{h}O(\tau',x')$ to the CFT action, where the external field $\hat{\ap}$ is proportional to $\ap$ when $\ap$
is small. In the presence of such a source we can estimate the one-point function by a perturbation with respect to the source as follows 
\ba
\la e^{\hat{\ap}\int d\tau' dx'\s{h}O(\tau',x')} O(\tau,x)\lb
& \simeq & \hat{\ap}\int d\tau' dx'\s{h} \la O(\tau',x')O(\tau,x)\lb \no
& = & \hat{\ap}\int \frac{d\tau' dx'}{x'^2}
\f{(x')^{\Delta}}{\left[(x-x')^2+(\tau-\tau')^2\right]^{\Delta}}\no
& \propto & \frac{1}{x^{\Delta}},
\ea
which indeed reproduces the general form (\ref{onepp}). Note that in the above calculation we took into the Weyl factor $e^{-\Delta\phi}=(x')^\Delta$ from the operator $O(\tau',x')$ inserted on $Q$ because the induced metric on $Q$ is given by 
$h_{\tau\tau}=h_{xx}=\frac{1}{(x')^2}$. In this way we can obtain the non-trivial one point function in a BCFT from the 2d gravity description.

\section{Conclusions and Discussions}
\label{sec:conclusions}
In this paper, we argued an equivalence relation, which we call Island/BCFT correspondence, between two theories: (i) a CFT in a right half plane coupled to an induced gravity in the left half one, and (ii) a BCFT, as in Figure~\ref{fig:Island}. We focused on the two dimensional case and 
confirmed this equivalence by examining the calculation of entanglement entropy, the boundary condition of energy stress tensor and one-point functions. We also identified the 2d induced gravity which is dual to the boundary dynamics of a 2d BCFT. 

The Island prescription of computing entanglement entropy in the theory (i) is equivalent to the mirror charge calculation of the field theoretic entanglement entropy in the theory (ii). The complete reflection boundary condition of energy flux $T_{xt}=0$ in (ii) is also obtained in (i) because the dynamical metric in 2d gives the vanishing of energy stress tensor, namely the Virasoro constraint. We derived the 2d induced gravity from its 3d gravity dual and found that it is given by a 2d CFT coupled to a 2d gravity whose action is simply given by the cosmological constant term. After integrating out the CFT fields, we obtain the Liouville action (\ref{LVef}) or more covariantly a specific type of dilaton gravity (\ref{LVeff}).
We also showed that to obtain non-vanishing bulk one-point functions in BCFTs (\ref{onepp}), we need to modify the original prescription of AdS/BCFT by considering a non-trivial background of matter fields. The Neumann-like boundary conditions of matter fields (\ref{EOWbca}) and (\ref{EOWbcaa}), imposed on the EOW brane, induce this non-trivial matter field background. We obtained analytical solutions for the calculation of a holographic one point function of an exactly marginal operator. For more general scalar operators we found numerical solutions. We confirmed that this new prescription correctly reproduces non-trivial one point functions. We also explained how we obtain the non-vanishing one point functions from the theory (i).  For this we noted that due to the background bulk scalar field in the gravity dual, a source is turned on for the operator dual to the scalar field and this leads to a non-trivial one point function.

There are several interesting future directions. An obvious one is a higher dimensional generalization, which we will come back soon \cite{higherD}.
Another important problem is to explore string theory embedding of the Island/BCFT correspondence and see how the coefficients of one-point functions behave in such top down models. It will also be intriguing to extend the AdS/BCFT construction and the Island/BCFT correspondence to gravity duals of more general critical theories \cite{Kachru:2008yh,Gouteraux:2011ce,Ogawa:2011bz,Huijse:2011ef,Dong:2012se}.

\section*{Acknowledgements}

We are grateful to Keisuke Izumi, Taishi Kawamoto, Takato Mori, Tatsuma Nishioka, Tetsuya Shiromizu, Yu-ki Suzuki, Norihiro Tanahashi, Tomonori Ugajin and Zixia Wei for useful discussions. We would also like to thank Kenya Ikeda for his collaboration at the early state of this work. This work is supported by the Simons Foundation through the ``It from Qubit'' collaboration and by MEXT KAKENHI Grant-in-Aid for Transformative Research Areas (A) through the ``Extreme Universe'' collaboration: Grant Number 21H05187.
TT is also supported by Inamori Research Institute for Science and World Premier International Research Center Initiative (WPI Initiative)
from the Japan Ministry of Education, Culture, Sports, Science and Technology (MEXT),
by JSPS Grant-in-Aid for Scientific Research (A) No.~21H04469 and
by JSPS Grant-in-Aid for Challenging Research (Exploratory) 18K18766.

\appendix
\section{Notation}
\label{app:notation}
In this paper, we use various coordinates to discuss the AdS$_3$/BCFT$_2$ and its braneworld holography. 
For readers convenience, we summarize our notation in this appendix.
When we combine whatever three-dimensional coordinates, we use indices $a, b, \cdots=1,2,3$,
while when we combine whatever two-dimensional coordinates, we use indices $i, j, \cdots=1,2$.
The bulk three-dimensional metric is denoted by $g_{ab}$, the induced metric on the brane is denoted by $h_{ij}$
and the induced metric on the asymptotic boundary is denoted by $\g_{ij}$.
We also use the following explicit coordinates.

$\bullet$ AdS$_3$ Poincare coordinates:
	\begin{align}
		ds_{\rm AdS_3}^2 \, &= \, \frac{-dt^2 + dy^2 +dz^2}{z^2} \, , \qquad ({\rm for}\ {\rm Lorentzian}) \\
		&= \, \frac{d\t^2 + dy^2 +dz^2}{z^2} \, , \qquad \ \ ({\rm for}\ {\rm Euclidean})
	\end{align}

$\bullet$ AdS$_2$ foliation coordinates of AdS$_3$:
	\begin{align}
		ds_{\rm AdS_3}^2 \, &= \, d\r^2 + \cosh^2\r \left( \frac{d\h^2 - dt^2}{\h^2} \right) \, , \qquad ({\rm for}\ {\rm Lorentzian}) \\
		&= \, d\r^2 + \cosh^2\r \left( \frac{d\h^2 + d\t^2}{\h^2} \right) \, , \qquad ({\rm for}\ {\rm Euclidean})
	\end{align}

$\bullet$ Janus AdS$_3$ coordinates:
	\begin{align}
		ds_{\rm Janus \ AdS_3}^2 \, &= \, d\r^2 + f(\r) \left( \frac{d\h^2 - dt^2}{\h^2} \right) \, , \qquad ({\rm for}\ {\rm Lorentzian}) \\
		&= \, d\r^2 + f(\r) \left( \frac{d\h^2 + d\t^2}{\h^2} \right) \, , \qquad ({\rm for}\ {\rm Euclidean})
	\end{align}

$\bullet$ AdS$_2$ coordinates on the brane:
	\begin{align}
		ds_{\rm AdS_2}^2 \, &= \, \frac{d\h^2 - dt^2}{\h^2} \, , \qquad ({\rm for}\ {\rm Lorentzian}) \\
		&= \, \frac{d\h^2 + d\t^2}{\h^2} \, , \qquad ({\rm for}\ {\rm Euclidean})
	\end{align}

$\bullet$ BCFT$_2$ coordinates:
	\begin{align}
		ds_{\rm BCFT_2}^2 \, &= \, -dt^2 + dy^2 \, , \qquad ({\rm for}\ {\rm Lorentzian}) \\
		&= \, d\t^2 + dy^2 \, , \qquad \ \ ({\rm for}\ {\rm Euclidean})
	\end{align}
where $y$ is restricted $-\d \le y$.\\

$\bullet$ Two-dimensional extended coordinates:
We combine the AdS$_2$ coordinates on the brane and the BCFT$_2$ coordinates by introducing a new spacial coordinate $x$ as
	\begin{align}
		ds_{(2)}^2 \, &= \, \frac{\d^2(dx^2 - dt^2)}{x^2} \quad (x<-\d) \, , \qquad
		ds_{(2)}^2 \, = \, -dt^2 + dx^2 \quad (-\d \le x) \qquad ({\rm for}\ {\rm Lorentzian}) \\
		ds_{(2)}^2 \, &= \, \frac{\d^2(dx^2 + d\t^2)}{x^2} \quad (x<-\d) \, , \qquad
		ds_{(2)}^2 \, = \, d\t^2 + dx^2 \quad (-\d \le x) \qquad \ ({\rm for}\ {\rm Euclidean}) 
	\end{align}

$\bullet$ Light cone coordinates of AdS$_2$:
	\begin{align}
		x^+ \, &= \, t + \h \, , \quad \ \, x^- \, = \, -t + \h \, , \qquad \ \, ({\rm for}\ {\rm Lorentzian}) \\
		x^+ \, &= \, i \t + \h \, , \quad x^- \, = \, - i \t + \h \, , \qquad ({\rm for}\ {\rm Euclidean})
	\end{align}
which gives
	\begin{align}
		ds_{\rm AdS_2}^2 \, = \, \frac{4dx^+dx^-}{(x^++x^-)^2} \, , \qquad  ({\rm for}\ \, {\rm both \ \, Lorentzian \ \, and \ \, Euclidean})
	\end{align}

\section{On-shell Action of Induced Gravity}
\label{app:onshell action}
In this appendix, we present a naive discussion for the derivation of the energy refrection eqaution in the higher-dimensional version of the AdS/BCFT set-up.
More complete discussion will be presented in \cite{higherD}.
We would like to compute the on-shell action of the induced gravity on the brane.
The total action we start with is given by
	\begin{align}
		I_{\rm total} \, = \, I_{\rm bulk} \, + \, I_{\rm brane} \, + \, I_{\rm bdy} \, ,
	\end{align}
where
	\begin{align}
		I_{\rm bulk} \, &= \, - \, \frac{1}{16\pi G} \int_M d^{d+1}x \, \sqrt{g}(R_{d+1}- 2\L) \, , \\
		I_{\rm brane} \, &= \, - \, \frac{1}{8\pi G} \int_Q d^dx \, \sqrt{h}(K_Q - T) \, , \\
		I_{\rm bdy} \, &= \, - \, \frac{1}{8\pi G} \int_\S d^dx \, \sqrt{\g} K_\S \, .
	\end{align}
As discussed in \cite{Boruch:2020wax,Boruch:2021hqs}, the induced gravity on the brane is described by 
	\begin{align}
	    I_{\rm bulk} \, + \, I_{\rm brane} \, \Rightarrow \, I_{\S} \, + \, I_{\rm IG} \, + \, I_{\rm ct} \, ,
	\end{align}
where
	\begin{align}
		I_{\S} \, &= \, - \, \frac{(d-1)V_{d-1} L}{\k^2 \e^d} \, , \\
		I_{\rm IG} \, &= \, \frac{(d-1)V_{d-1}}{\k^2} \int_{-\inf}^0 dx \, e^{d\p} \left[ \sqrt{1-e^{-2\p} \dot{\p}^2} \, + \, e^{-\p} \dot{\p} \arcsin \big( \dot{\p} e^{-\p} \big) \, + \, \frac{T}{d-1}\right] \, , \\
		I_{\rm ct} \, &= \, - \, \frac{V_{d-1}}{\k^2} \bigg[ e^{(d-1)\p} \arcsin \big( \dot{\p} e^{-\p} \big) \bigg]^0_{-\inf} \, ,
	\end{align}
where the dot denotes a derivative with respect to $x$. We also defined $L$ as the length of the $y$ direction on the boundary and $V_{d-1}$ as the rest of $d-1$ dimensional volume.
The constant contribution $I_{\S}$ is the contribution from the BCFT on $\S$ with width given by $L-\l \e$, where $\l$ is defined in (\ref{lambda}).
The background solution obtained from $I_{\rm IG}$ is given by 
	\begin{align}
	    e^{2\p} \, = \, \frac{1}{\Big( -\sqrt{1-(\frac{T}{d-1})^2} \, x + \e \Big)^2} \, , 
	\end{align}
and one can compute the on-shell actions for the induced gravity as
	\begin{align}
		I_{\rm IG}^{\onshell} \, &= \, \frac{V_x}{\k^2} \frac{\arcsin\sqrt{1- (\frac{T}{d-1})^2}}{\e^{d-1}} \, , \\
		I_{\rm ct}^{\onshell} \, &= \, - \, \frac{V_x}{\k^2} \frac{\arcsin\sqrt{1- (\frac{T}{d-1})^2}}{\e^{d-1}} \, .
	\end{align}
Therefore, the total on-shell action vanishes due to the precise cancellation between $I_{\rm IG}^{\onshell}$ and the counterterm.
This implies that the ADM energy of the induced gravity also precisely zero. 
Hence the energy flux equation between the induced gravity on the brane and the boundary CFT on the asymptotic boundary $\S$ (which we will describe more in detail in next section) gives 
	\begin{align}
		T_{x^+x^+} \, - \, T_{x^-x^-} \, = \, 0 \, ,
	\end{align}
for any dimension.
Here $x^{\pm}$ are the light cone coordinates constructed by the spacial normal coordinate of the interface and time direction.

However this discussion is naive as we did not care appropriately the cutoff $\e$.
We will come back to this question in the near future \cite{higherD}.

\section{ADM Energy in JT Gravity}
\label{app:adm energy}
This appendix is a quick summary of the ADM Energy in JT gravity. For more complete discussion, refer to \cite{Maldacena:2016upp}.

For two-dimensional gravity, the ADM energy is defined by (e.g. see \cite{Grumiller:2002nm})
	\begin{align}
		8\pi G M \, = \, \sqrt{h} (-\pa_n \p + \p) \, ,
	\end{align}
where $h$ is the induced metric on the boundary and $\pa_n$ denotes the derivative with respect to an outward pointing unit normal vector.
In JT gravity, the boundary surface is defined by ($\t(u), \h(u)$), where $u$ is a parameter called ``boundary time'' \cite{Maldacena:2016upp}.
The boundary condition of the metric fixes $\h(u) = \e \t'(u) + \cdots$, while the boundary condition of dilaton gives $\p|_{bdy} = \bar{\p}_r/\e$.
Therefore, the normal derivative is now given by
	\begin{align}
	    \pa_n \, = \, \e (\h' \pa_\t - \t' \pa_\h) \, ,
	\end{align}
and the induced metric is $\sqrt{h} = \e^{-1}$. Combining these, the ADM energy is now written as
	\begin{align}
		8\pi G M \, = \, \t' \pa_\h \p \, - \, \h' \pa_\t \p \, + \, \frac{\p}{\e} \, .
	\label{ADM1}
	\end{align}
Using the $(\h, \h)$ component of the dilaton equation
	\begin{align}
		\pa_\t^2 \p \, - \, \frac{1}{\h} \pa_\h \p \, - \, \frac{1}{\h^2} \, \p \, = 0 \, ,
	\end{align}
we can eliminate the $\pa_\h \p$ term from the RHS of (\ref{ADM1}).
Then, rewriting $\pa_\t = (\t')^{-1} \pa_u$, we find the ADM energy is given by the Schwarzian derivative 
	\begin{align}
		8\pi G M \, = \, \bar{\p}_r \{ \t(u) , u \} \, .
	\end{align}

\section{Replica Wormholes in JT Gravity}
\label{app:replica wormholes}
In this appendix, we review the replica wormholes in the system of JT gravity coupled to conformal matter fields \cite{Almheiri:2019psf}.
In JT gravity the energy flux equation is given by (\ref{JT_energy_flux})
	\begin{align}
		\frac{\bar{\p}_r}{8\pi G} \, \pa_u \{\t, u \} \, = \, - i (\t')^2 \big( T_{x^+x^+} - T_{x^-x^-} \big) \, .
	\end{align}
If we set $\t = e^{2\pi i u/\b}$, then we have 
	\begin{align}
		T_{yy} - T_{\bar{y}\bar{y}} \, &= \, \left( \frac{d\t}{du} \right)^2 \big( T_{x^+x^+} - T_{x^-x^-} \big)
		\, = \, - \left( \frac{2\pi}{\b} \right)^2 \big( T_{x^+x^+} - T_{x^-x^-} \big) \, ,
	\end{align}
so the energy flux equation is written as
	\begin{align}
		\frac{\bar{\p}_r}{8\pi G} \left( \frac{\b}{2\pi} \right)^2 \, \pa_u \{\t, u \} \, = \, i \big( T_{yy} - T_{\bar{y}\bar{y}} \big) \, .
	\end{align}

For replicated geometry, we have 
	\begin{align}
		\tilde{\t} \, = \, \left( \frac{\t-A}{1-A \t}\right)^{\frac{1}{n}} \, ,
	\end{align}
so the energy flux equation becomes
	\begin{align}
		\frac{\bar{\p}_r}{8\pi G} \left( \frac{\b}{2\pi} \right)^2 \, \pa_u \{\tilde{\t}, u \} \, = \, i \big( T_{yy} - T_{\bar{y}\bar{y}} \big) \, .
	\end{align}
We can easily evaluate the Schwarzian derivative and find
	\begin{align}
		\{\tilde{\t}, u \} \, = \, \frac{2\pi^2}{\b^2} \, - \, \frac{2\pi^2}{\b^2} \left( 1 - \frac{1}{n^2} \right) \frac{(1-A^2)^2}{|1-A \t|^4} \, .
	\end{align}
The transformation of the stress tensor is evaluated as in section~\ref{sec:replica wormholes}.
Therefore, in the replicated geometry, the energy flux equation is found as
	\begin{align}
		\pa_u R(u) \, = \, \frac{cG}{3\bar{\p}_r} \, \mathcal{F}(e^{\frac{2\pi i u}{\b}}) \, ,
	\end{align}
where $\mathcal{F}$ is defined in (\ref{mathcalF}) and 
	\begin{align}
		R(u) \, = \, \frac{(1-A^2)^2}{|1-A \t|^4} \, .
	\end{align}
Finally Fourier transforming from $\t$ to k, the $k=1$ equation gives
	\begin{align}
		0 \, &= \, \int_0^\b du \, e^{-\frac{2\pi i u}{\b}} \left( \frac{cG}{3\bar{\p}_r} \, \mathcal{F} \, - \, \pa_u R(u) \right) \nn\\
		&= \, \frac{\b cG}{3\pi \bar{\p}_r} \, \frac{\sinh\frac{\pi(a-l)}{\b}}{\sinh\frac{\pi(a+l)}{\b}} \, - \, \frac{1}{\sinh\frac{2\pi a}{\b}} \, .
	\end{align}
This equation gives the same condition as derived from the quantum extremal surface:
	\begin{align}
		\sinh\frac{2\pi a}{\b} \, = \, \frac{3\pi \bar{\p}_r}{\b cG} \, \frac{\sinh\frac{\pi(a+l)}{\b}}{\sinh\frac{\pi(a-l)}{\b}} \, .
	\end{align}

\bibliographystyle{JHEP}
\bibliography{IsBCFT}


\end{document}